\pdfoutput=1
\documentclass[aps,pra,showpacs,twocolumn,nofootinbib,floatfix]{revtex4}
\usepackage{amsmath,graphicx,epsfig,amssymb,subfigure}
\usepackage[usenames]{color}

\newcommand{\h}[2][ ]{\hat{#2}^{\vphantom{\dag} #1}}
\newcommand{\hd}[2][ ]{\hat{#2}^{\dag #1}}

\begin{document}

\title{Dynamical instabilities of Bose-Einstein condensates at the band-edge
in one-dimensional optical lattices}

\author{Andrew~J.~Ferris}
\author{Matthew~J.~Davis}
\affiliation{ARC Centre of Excellence for Quantum-Atom Optics,
School of Physical Sciences, University of
Queensland, Brisbane QLD 4072, Australia.}

\author{Reece~W.~Geursen}
\author{P.~Blair Blakie}
\author{Andrew~C.~Wilson}
\affiliation{Jack Dodd Centre for Photonics and Ultra-Cold Atoms,
Department of Physics, University of Otago, P.O. Box 56, Dunedin,
New Zealand.}

\date{\today}

\begin{abstract}

We report on experiments that demonstrate dynamical instability in a
Bose-Einstein condensate at the band-edge of a one-dimensional optical lattice.
The instability manifests as rapid depletion of the condensate and conversion to
a thermal cloud. We consider the collisional processes that can occur in such a
system, and perform numerical modeling of the experiments using  both a
mean-field and beyond mean-field approach. We compare our numerical results to
the experimental data, and find that the Gross-Pitaevskii equation is not able
to describe this experiment.  Our beyond mean-field approach, known as the
truncated Wigner method, allows us to make quantitative predictions for the
processes of parametric growth and thermalization that are observed in the
laboratory, and we find good agreement with the experimental results.

\end{abstract}


\maketitle

\section{Introduction}

Since the realization of dilute-gas Bose-Einstein condensates (BECs)
\cite{Anderson1995a,KBDavis1995a,Bradley1995a} there have been many
experiments on BECs in one-dimensional (1D) optical lattices
\cite{Ovchinnikov1999,Orzel2001,Hensinger2001,
Cataliotti2001,Greiner2001,
Cristiani2002,HeckerDenschlag2002,Mellish2003a,Eiermann2004,
WuNiuBattle, Cataliotti2003a, Cristiani2004a, Fallani2004a,
DeSarlo2005a,Fertig2005,Katz2004a,Katz2005a}. Neutral atoms in
periodic potentials demonstrate a wide range of phenomena that are
familiar from other physical systems. Dilute gas BECs are attractive
systems in which to study these effects due to the flexibility and
control of the experimental apparatus, combined with the existence
of microscopic theories that are computationally tractable with
various degrees of approximation.

At the simplest level, atoms in an optical lattice are described by
Bloch waves familiar from the description of electrons in periodic
condensed matter systems. At sufficiently large densities the
interatomic interactions result in a rich range of nonlinear
phenomena. Notably, the quantum phase transition from a superfluid
to a Mott-insulator \cite{Greiner2002, Orzel2001} and generation of
bright gap solitons \cite{Eiermann2004} have been observed. When the
condensate is moving relative to the lattice, the nonlinearity can
induce a dynamical (or modulational) instability that leads to an
observed thermalization of the condensate \cite{WuNiuBattle,
Cataliotti2003a, Cristiani2004a, Fallani2004a,
DeSarlo2005a,Fertig2005,Katz2004a,Katz2005a}.

There has been a large amount of theoretical work devoted to
instabilities in optical lattices
\cite{Katz2004a,Katz2005a,Abdullaev2001a, Konotop2002a, Wu2003a,
Menotti2003a, Machholm2003a,Scott2003a, Machholm2004a, Zheng2004a,
Modugno2004a, Nesi2004a, Modugno2005a,Iigaya2006a,Polkovnikov2004a,
Ruostekoski2005a, Polkovnikov2005a, GeaBanacloche2006a, Rey2005a,
Ponomarev2006a, Altman2005a}. Two types
of instabilities are known to occur: dynamical and energetic.
Energetic (or Landau) instabilities exist when the superfluid or
condensate is not at a local minimum of energy, and can deplete the
condensate via dissipative processes such as interactions with a
thermal cloud \cite{Modugno2005a,Iigaya2006a}. Dynamical
instabilities manifest as a exponential growth of certain modes and
are related to the process of parametric amplification. A dynamical
instability will rapidly deplete even a pure, isolated condensate.
This process will often be chaotic and cause period doubling
\cite{Machholm2004a, Gemelke2005a}, turbulence and vortex formation
\cite{Scott2003a}, and a loss of coherence of the condensate. A
dynamically unstable BEC will eventually thermalize to a higher
temperature.

On the other hand, the parametric amplification that causes the dynamical
instability can be used to generate interesting states of the condensate.
Hilligs{\o}e and M{\o}lmer~\cite{Hilligsoe2005a,Molmer2006a} have
shown that momentum and energy can be
conserved in non-trivial collisions between atoms in a 1D system
with a periodic potential. This is in contrast to free space, where such
collisions are energetically forbidden. This process is analogous to
phase-matching in optical parametric amplification, and has been observed with
a Bose-Einstein condensate both spontaneously and with seeding by
Campbell \emph{et al}.~\cite{Campbell2006a}. %Seeding significantly
%increases the rate of collisions due to the Bose-enhanced nature of the
%process.
Related parametric amplification processes have been observed in systems with
modulated lattices \cite{Gemelke2005a}. Parametric generation creates
sub-Poissonian number correlations and quadrature entanglement between
different modes in optical experiments, and is expected to do the same for
atomic systems \cite{Olsen2006a}.

To quantitatively simulate dynamical instabilities in a BEC it is
necessary to use beyond-mean-field method for quantum dynamics.  In
this paper we will be making use of the truncated Wigner
method~\cite{steele1998, gardiner1999a, Gardiner2003}. This method
has successfully been previously applied to condensates in optical
lattices by a number of authors. Firstly, Polkovnikov and Wang
studied the unstable dynamics of an condensate offset in a combined
optical lattice plus harmonic potential~\cite{Polkovnikov2004a},
which lead to dampened Bloch oscillations. Isella and Ruostekoski
investigated the dynamics of a trapped condensate as an optical
lattice was switched on and found number squeezing in experiments is
limited because adiabatic ramping takes a very long
time~\cite{Isella2005a,Isella2006a}.  In another work the same authors
found that quantum fluctuations cause dissipation via the dynamical instability in the
system~\cite{Ruostekoski2005a}. Katz~\emph{et al}. investigated
scattering due to the dynamical instability at the band-edge of an
optical lattice using a 2D model~\cite{Katz2005a}. They observe that
the structure of the scattering halo can deviate strongly from the
s-wave halo in free space.

In this paper we report on experiments where we observe rapid
thermalization of a Bose-Einstein condensate at the band-edge of a
one-dimensional optical lattice. We then  model our system using a
mean-field approach, but demonstrate that we are unable to
accurately describe the dynamics of our system with the
Gross-Pitaevskii equation. We then employ the truncated Wigner
method  to simulate
beyond mean-field quantum dynamics of the system starting from a
coherent condensate at zero temperature.  In our simulations, we observe dynamical instabilities
that lead to heating in qualitative agreement with the experimental
results. We attempt quantitative modeling of the experiments and
compare our numerical results to the experimental data.

This paper is organized as follows.  In Sec.~\ref{sec_experiment} we
describe our experiments where a BEC is loaded into a 1D optical
lattice and our observations of the resulting dynamics.
Section~\ref{sec_mean_field} models the experiments using the
mean-field Gross-Pitaevskii equation, and describes the physics that
can be captured by this approach. In Sec.~\ref{sec_dyn_inst} we
describe the nature of the dynamical instability in various
parameter regimes including those relevant to our experiments. We
introduce our quantum model of the system in Sec. \ref{sec_model},
and discuss its advantages and limitations.
Section~\ref{sec_tw_results} describes our numerical results and
studies the instabilities present for systems of different
dimensionality. In Sec.~\ref{sec_thermal} we discuss important
experimental considerations that have not been implemented in our
simulations and use our results to develop a quantitative estimate
of the heating rates that are observed in the experiments, before
concluding in Sec.~\ref{sec_conclusion}.

%Finally, in Sec. \ref{sec_6} we discuss
%the regimes that the dynamical instability (\todo I think all this could go in
%section 3).

%\cite{Deng1999a} This paper does only the stimulated part of
%\cite{Vogels2002a}. Good for intro?

\section{Experimental results}

\label{sec_experiment}

\subsection{Experimental Set-Up}

Our experimental set-up has previously been described by Mellish \textit{et
al}.~\cite{Mellish2003a}. Briefly, we use a double magnetic-optical
trap set-up to collect and pre-cool $^{87}$Rb atoms.  The sample is
then loaded into a QUIC magnetic trap, from which evaporative
cooling results in a Bose-Einstein condensate of approximately $1
\times 10^5$ atoms in the $|F\!\!=\!\!2,m_F\!\!=\!\!2\rangle$
hyperfine ground state with a condensate fraction of approximately
80\%. The peak density in the magnetic trap is $\sim 1.7 \times
10^{20}$ m$^{-3}$ with an axial trap angular frequency of $\omega_x
\approx 2\pi \times 14.6$ Hz and a radial angular frequency of
$\omega_y = \omega_z \approx 2\pi \times 179$ Hz.
%\color{green} These are different from Reece's thesis which says $\omega_r =
%9\omega_z = 2\pi \times 162.3$ Hz. Any comments? \color{black}

In our experiments a weak  moving optical lattice is suddenly
applied to the condensate at an angle of 63~$\pm$~$3^{\circ}$ to the
weak axis of the trap. The lattice potential is generated from a
single laser of wavelength 780 nm detuned 4.49 GHz from atomic
resonance to minimize spontaneous emission. In the lab frame, the
two lasers of wave-number $|\mathbf{k}_L| = k_L \approx
2\pi/(780\;\mathrm{nm})$ and angular frequency difference $\delta$
produce a moving sinusoidal potential
\begin{equation}
  V_{L}(\mathbf{r},t) = \frac{V_0}{2} \cos\left(2 \mathbf{k}_L \cdot
\mathbf{r} - \delta t\right) ,
  \label{potential}
\end{equation}
where $V_0$ depends on the intensity of the lasers and the
interaction with the atoms. For ease of analysis we define:
\begin{equation}
  V_0 = s \frac{\hbar^2 k_L^2}{2m},
\end{equation}
where $m$ is the mass of the atoms, and $s$ characterizes the
strength of the lattice. The frequency difference between the beams
is chosen so that relative to the lattice the atoms are moving with
a quasi-momentum corresponding to the Brillioun zone edge (the Bragg
condition). In the frame of the lattice, the atoms are moving with
momentum $-\hbar \mathbf{k}_L$.

\subsection{Bragg Scattering}

We can gain a basic understanding of the system in the dilute,
collisionless limit. The single-particle eigenfunctions of the
Hamiltonian without the magnetic trap are the spatially periodic
Bloch states that are superpositions of momentum components
differing by $2\hbar k_L$. Diagonalizing this in the limit $s \ll 1$
one finds the Bloch states at the edges of the two lowest bands are
superpositions of the two momentum states $|\!\pm\!k_L\rangle$ (in
the frame of the lattice). Denoting the Bloch state with
quasi-momentum $\hbar q$ in the $n$th band by $|q,n\rangle$, we have
the result:
\begin{eqnarray}
   |k_L, 1 \rangle (t) & = & \frac{|\!+\!k_L\rangle - |\!-\!k_L \rangle}{\sqrt{2}}e^{-i\omega_1 t} \nonumber \\
   |k_L, 2 \rangle (t) & = & \frac{|\!+\!k_L\rangle + |\!-\!k_L \rangle}{\sqrt{2}}e^{-i\omega_2
   t}, \label{BlochStates}
\end{eqnarray}
and that $\hbar\omega_2 - \hbar\omega_1 = V_0/2$. This means that an
atom suddenly placed in the lattice with definite momentum $+k_L$ is
in a superposition of the above modes, and furthermore the
momentum will Rabi oscillate between $\pm k_L$ at a frequency
$\omega_2 - \omega_1$ in a process rather similar to a Bloch
oscillation.

By applying the lattice for an appropriate amount of time one can
transfer any fraction of the condensate from one
momentum state to the other (Bragg scattering). Alternatively, by phase
shifting the optical lattice at the appropriate time one can
non-adiabatically load a weakly-interacting condensate into an
eigenstate of the lattice at either of the lowest two band edges, as
described previously in Mellish \textit{et al}.~\cite{Mellish2003a}.

\subsection{Experiments in the Collisionless Regime}
\label{sec:linear_expt}

\begin{figure}[t]
  \includegraphics[width=8.6cm]{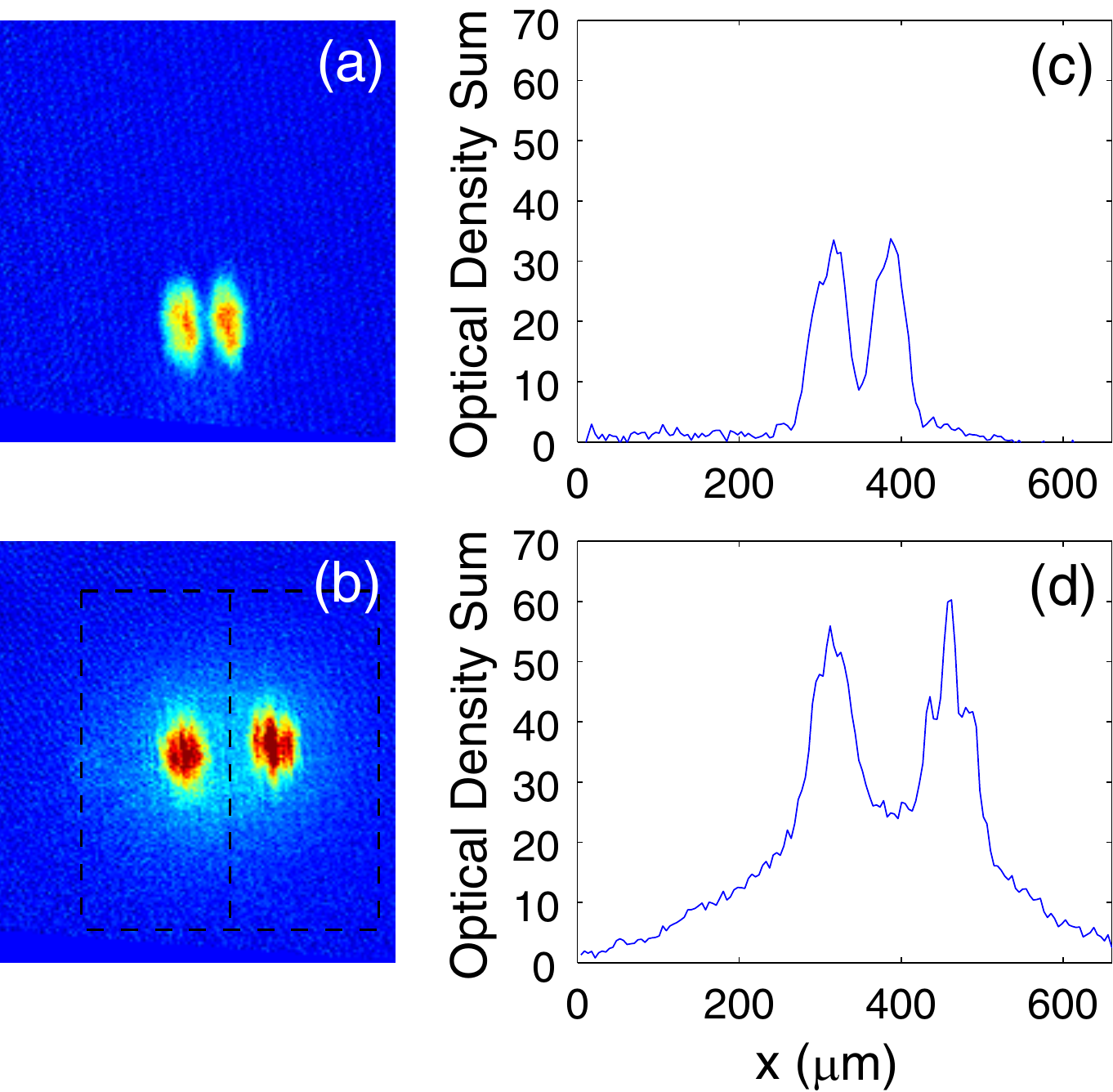}
  \caption{Typical time-of-flight images of a condensate after it has been in an
optical lattice at the Brillouin zone boundary (Rabi cycling).
Atomic clouds to the right are the atoms that have been Bragg
scattered (excited). (a) Collisionless regime, $t$~=~280~$\mu$s
(applied to freely expanding atoms 10~ms after being released from the trap),
after 23~ms total time of flight.
(b) Nonlinear regime, $t$~=~240~$\mu$s (applied to trapped atoms)
and 29~ms time of flight. The dashed boxes show typical regions of
interest used for fitting the data. Also, the clouds in (b) are
further apart and larger than those in (a) due to an additional
16~ms of expansion after the lattice is removed. (c) and (d) are
$x$-axis profiles of the absorption images (a) and (b) respectively,
summed along the vertical axis. %Note the x axis is a relative scale with a different origin
%with respect to the overall CCD array for each plot.
\label{fig_exp_photo}}
\end{figure}

\begin{figure}[t]
  \includegraphics[width=7cm]{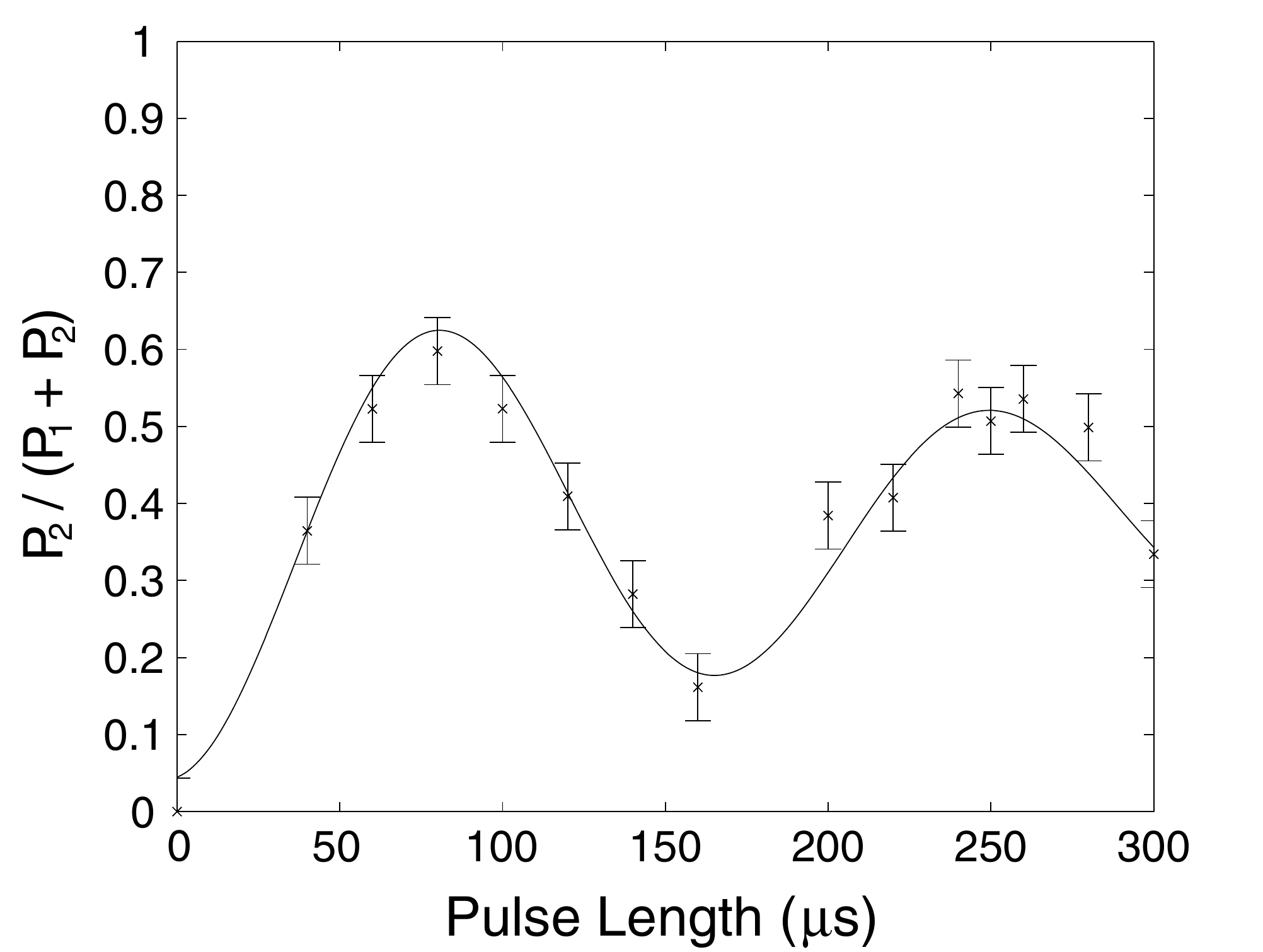}
  \caption{Fraction of the population that has momentum $> 0$ in frame of the lattice (i.e. the Bragg scattered atoms) as a function of the amount of time the lattice is applied after 10 ms of free expansion. \label{fig_exp_1}}
\end{figure}

We present two sets of experiments in this paper.  In the first,
after creating the condensate we turn off the magnetic trap and
allow the condensate to expand for 10~ms in order to reduce the
effect of interactions.  This allows for the dissipation of
essentially all of the mean-field interaction energy.
 We then suddenly turn on and hold the optical lattice potential
for a total time $t$, followed by a further 13~ms expansion (totalling 23~ms
since being released from the trap) before imaging. In the frame of
the lattice, the condensate begins with momentum $-\hbar k_L$, and
as expected we find that the condensate undergoes Rabi oscillations
between $\pm\hbar k_L$ momentum in this frame.

An absorption image after a lattice evolution time of
$t$~=~280~$\mu$s is provided in Fig.~\ref{fig_exp_photo}(a), where
we see roughly equal populations in the two momentum modes. To find
the populations in each mode, we perform a bimodal fit to the data
at momentum $+\hbar k_0$ and $-\hbar k_0$. In the region surrounding
each momentum state we fit a Gaussian distribution for the thermal
component and an inverted parabola for the condensate. We denote the
condensed population in the region of interest surrounding the
original condensate as $N_1$ and that in the second region of
interest as $N_2$. Similarly we define the total population
(condensed and non-condensed) in each region as $P_1$ and $P_2$.

The oscillation between the momentum states is plotted in
Fig.~\ref{fig_exp_1}. From the frequency of this Rabi oscillation
(5.9~$\pm$~0.6~kHz) we estimate the strength of our optical lattice,
and find that $s \approx 3.1 \pm 0.3$. The Rabi oscillation is
damped due to the finite momentum width of the wave function of the
initial condensate. An exponentially decaying sinusoid is fitted to
the data in Fig.~\ref{fig_exp_1} for which the damping time is
$400$~$\pm$~$200$~$\mu$s. We investigate the damping due to
dephasing theoretically in Sec.~\ref{sec_linear_dephasing}.

\subsection{Experiments in the Nonlinear Regime}
\label{sec:nonlinear_expt}

\begin{figure}[t]
  \includegraphics[width=8.5cm]{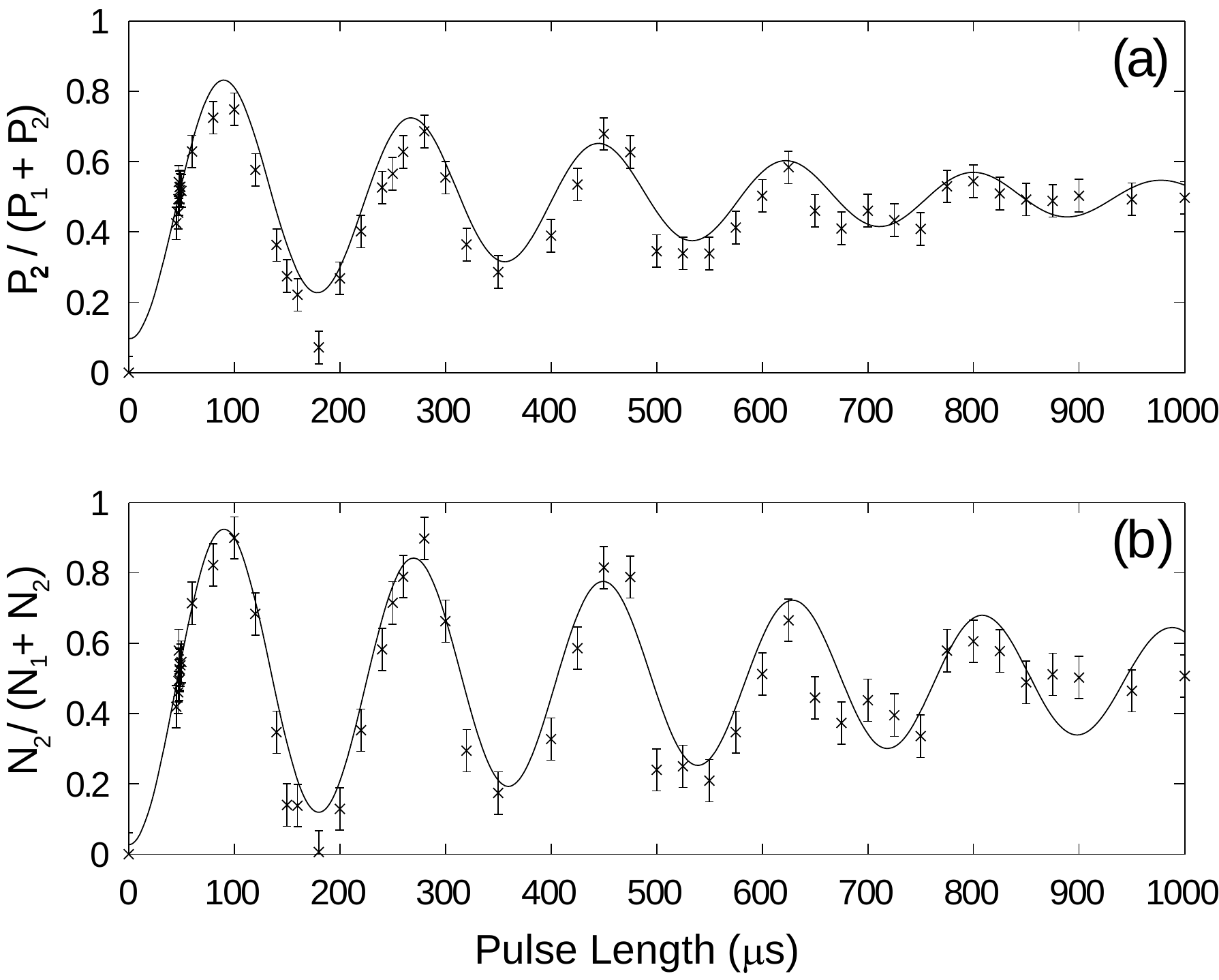}
  \caption{ In (a) we plot the fraction of the total population that has been Bragg scattered as a function of the time the lattice is switched on when the atoms are inside the
  trap. In (b) we plot the fraction of the condensed atoms, found with a Gaussian fitting procedure, that have been Bragg scattered. \label{fig_exp_2}}
\end{figure}

\begin{figure}[t]
  \includegraphics[width=7.2cm]{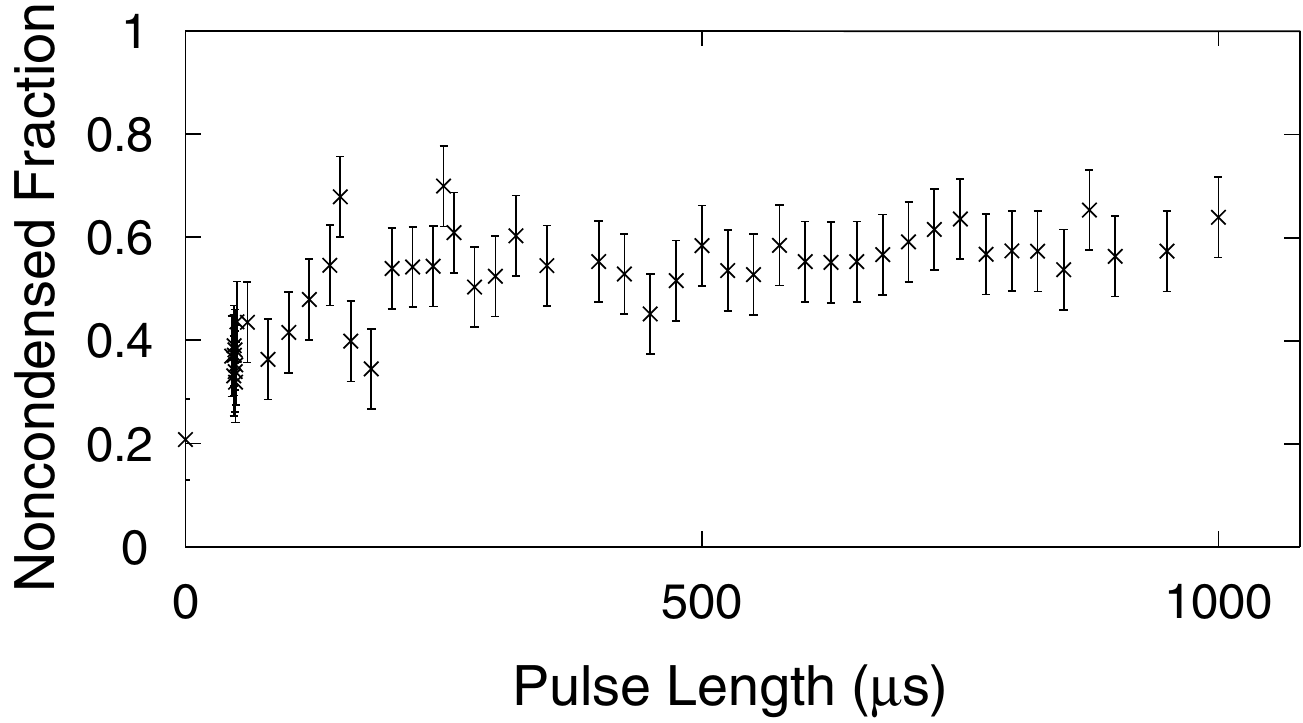}
  \caption{The non-condensed fraction is plotted as a function of the time the lattice is left on inside the trap.%: (a)
  %the lattice is not phase shifted%; (b) the lattice is phase shifted such that the atoms are loaded into the lowest band-edge Bloch state
 \label{fig_noncondensed}}
\end{figure}

In the second set of experiments the optical lattice is suddenly
turned on while the condensate is still confined in the magnetic
potential. The BEC evolves in the combined potential for a time $t$
before the lattice, and then the magnetic trap, are both rapidly
switched off.  An absorption image is then taken after 29~ms of free
expansion to observe the resulting momentum distribution.  For the
short range of times $t$ in these experiments there is essentially
no change in the atomic density during the evolution in the trap,
however as can be seen in Fig.~\ref{fig_exp_photo} the momentum
distribution undergoes significant evolution. The broad momentum
distribution indicates that atoms have been scattered to a variety
of different momentum states. The major difference between these
experiments and those described in Sec.~\ref{sec:linear_expt}
 is that the density
is much greater while the optical lattice is applied, suggesting that this
effect is caused by nonlinear processes.

We observe that a fraction of the atoms still undergo Rabi
oscillation between the $\pm \hbar k_L$ momentum modes. In
Fig.~\ref{fig_exp_2} we plot the oscillating populations of both the
entire sample, and the condensed modes found by the bimodal fitting
procedure, finding qualitative agreement with previous results in
\cite{Katz2004a}. By fitting an exponentially decaying sinusoid, we
observe a damping time of $\tau_P$~=~450~$\pm$~80~$\mu$s for the
entire sample. However, we find that the condensed modes oscillate
for much longer with a damping time of
$\tau_N$~=~800~$\pm$~100~$\mu$s. The fact that this is longer than
the damping time measured in the experiments in the collisionless
regime is due to the reduced momentum width of the condensate inside
the trap.  In the expansion process the interaction energy of the
condensate is converted to kinetic energy, resulting in a broader
momentum distribution.

The fraction of non-condensed atoms as a function of the evolution time in the
lattice is plotted in Fig.~\ref{fig_noncondensed}. One can see that the
non-condensed fraction grows rapidly until it saturates at slightly above 50\%.
This rapid generation of the thermal component has been termed `anomalous
heating' by other authors \cite{Dabrowska2006a}.  Understanding  this apparently
nonlinear process
is the main goal of this paper.

\section{Mean field dynamics}
\label{sec_mean_field}

In this section we turn to a common tool of choice for modelling BEC experiments
near $T=0$: the Gross-Pitaevskii equation (GPE).  The mean-field dynamics
described by the GPE have often proven to be in excellent agreement with
observations in a number of experiments \cite{Dalfovo1999a}.

\subsection{The Gross-Pitaevskii Model}

The time-dependent Gross-Pitaevskii equation for the evolution of
the macroscopic wave function $\psi(\mathbf{r},t)$ of a BEC is
\begin{equation}
  i \hbar \frac{\partial\psi(\mathbf{r})}{\partial t} =
 -\frac{\hbar^2}{2m}\nabla^2\psi(\mathbf{r}) + V(\mathbf{r}) \psi(\mathbf{r})
  + U_0 |\psi(\mathbf{r})^2|
 \psi(\mathbf{r}). \label{GPE}
\end{equation}
The characteristic strength of atomic interactions is
$U_0=4\pi\hbar^2 a_s / m$ where $a_s$  is the s-wave scattering
length which for $^{87}$Rb we take to be $a_s = 100$ Bohr radii.
The external potential $V(\mathbf{r},t)$ in the experiment is
\begin{eqnarray}
V(\mathbf{r},t) = \frac{m}{2}(\omega_x^2 x^2 + \omega_y^2 y^2 +
\omega_z^2 z^2) + V_{L}(\mathbf{r},t), \label{fullpotential}
\end{eqnarray}
which is the sum of the parabolic magnetic potential of the trap and
the sinusoidal optical potential from Eq.~(\ref{potential}). The
trap frequencies are those measured by the experiment:
$(\omega_x, \omega_y, \omega_z) = 2\pi \times (14.6,179,179)$ Hz.

Our simulations begin with $10^5$ atoms in the Thomas-Fermi ground
state, which has a peak density of $n_0$~$\approx$~$1.69 \times
10^{20}$ m$^{-3}$.

\subsection{Linear Regime} \label{sec_linear_dephasing}

\begin{figure}[t]
\begin{centering}
\includegraphics[width=7.5cm]{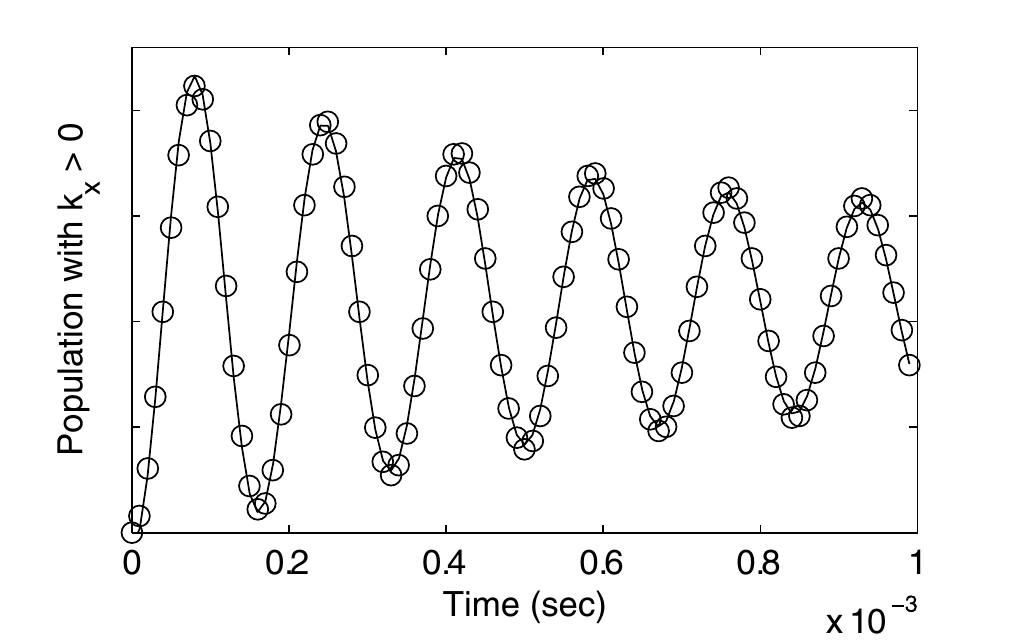}
\caption{ Comparison of our analytic estimate for the damped Rabi
oscillations due to dephasing of an non-interacting condensate with
a full numerical calculation. In both cases we have assumed a
Gaussian distribution of initial
  momenta which dephase with each other. Circles represent data from
  numerical simulations, and the line is the analytic result in
  Equation \ref{linearfitting}. } \label{linearrabi}
\end{centering}
\end{figure}

The density of the system decreases rapidly after release from the
trap, and so interactions between the particles become negligible in
the experiments where the lattice is applied after 10 ms of free
expansion. A simple one-body theory can therefore be applied to this
system [taking $U_0|\psi^2| = 0$ in Eq.~(\ref{GPE})]. We will now
derive an analytic approximation to the form of the damped
oscillations.
The Rabi frequency of oscillation of an atom is dependent on
its initial momentum, and the momentum width of the condensate gives
a spread of Rabi frequencies whose sum results in the
damping of the full oscillations.

It is worth noting that the dimensionality of the problem can be
reduced. Any momentum orthogonal to the lattice does not affect the
oscillation frequency, and therefore is not a part of our dephasing
analysis.  Thus, we only need to look at the distribution of momenta in the
lattice direction. The distribution is a result of the initial
momentum within the trap plus the released interaction energy and
would be difficult to calculate precisely. We introduce a few
approximations below.

To simplify analysis, our initial distribution was taken to be
Gaussian
\begin{equation}
 P(k) \propto \exp\left(\frac{(k - k_L)^2}{2\sigma_k^2}\right),
\label{linearinit}
\end{equation}
where almost all of our kinetic energy comes from the released
interaction energy $U_{\mathrm{int}} = \frac{2}{7}U_0 n_0$. Due to
the high aspect ratio of the trap, almost all this energy is
distributed equally in the two tightly-trapped dimensions.
 Given that the lattice is at angle $\theta \approx
63^{\circ}$ to the weak axis of the trap, the conservation of energy requires
that
\begin{equation}
  \frac{\hbar^2 \sigma_k^2}{2m} = \frac{\sin^2\theta}{7} U_0 n_0.
\end{equation}
This results in an predicted $\sigma_k \approx 1.23 \times 10^6$ m$^{-1}$.

We will now analyze the dynamics that these modes will undergo. In a
periodic potential, the eigenstates are superpositions of momentum
states corresponding to the same quasi-momentum (i.e.\ $\hbar k \pm 2 n \hbar k_L$
where $n$ is an integer).
We can then find the rate of oscillation between two modes of the
same quasi-momentum. At the band edge, we assume the eigenstates to
be superpositions of only the momentum modes $|\!\pm\!k_L\rangle$.
%and $|\!\pm\!3k_L \rangle$
The energy difference $\hbar \omega_0$ between the two states at the
band-edge
\begin{equation}
  \hbar \omega_0 = \frac{\hbar^2 k_L^2}{2 m} %\left(
  \frac{s}{2}, %+ \sqrt{16 + s + \frac{5 s^2}{64}} - \sqrt{16 - s
%  + \frac{5s^2}{64}}\right),
\end{equation}
is accurate for small values of $s$, and has a 1\% error (compared
to direct diagonalization) for the value of $s = 3.1$ used in these
simulations and experiments. The Rabi frequency of the oscillation
is $\omega_0$.

Next we will find how the oscillation frequency varies with the
quasi-momentum. We will assume that the eigenstates are
superpositions of  the modes $|\Delta\!k + k_L\rangle$ and
$|\Delta\!k - k_L\rangle$ only. In this basis, we see the energy
difference between the upper and lower states is (to second order in
$\Delta\!k$):
\begin{equation}
  \hbar \omega \approx \hbar \omega_0 + \frac{8 \hbar^2
  (\Delta\!k)^2}{sm} \label{secondorderapprox}.
\end{equation}
We can now use this dispersion relation, and the initial conditions of
Eq,~(\ref{linearinit}) to find the distribution as a function of
$\omega$. Performing the Fourier transform yields the population of
the negative momentum modes in time
\begin{equation}
  N^{(-)} = N_0^{(-)} \left( 1 - \frac{\cos\left[\omega_0 t +
  \arctan(t/\tau)/2\right]}{\left(1 +
  t^2/\tau^2\right)^{1/4}} \right) ,
  \label{linearfitting}
\end{equation}
where $\tau = sm/16 \hbar \sigma_k^2 \approx 174$ $\mu$s. Note that
the functional form of this decay is much slower than exponential, and goes
as $t^{-1/2}$ for $t\gg\tau$. To give an idea of the validity of the approximation
in Eq.~(\ref{secondorderapprox}), we also performed a numerical
simulation using the linear Schr\"{o}dinger equation and the same
Gaussian initial conditions. Fig.~\ref{linearrabi} demonstrates
agreement between the analytical estimate and the numerical results.

To compare with the experiment, we fitted an exponential decaying
sinusoid to the first 300 $\mu$s of data in Fig.~\ref{linearrabi},
yielding a decay time of 954 $\mu$s, compared to a measured time of
$400 \pm 200$ $\mu$s.

There are several reasons that might explain the discrepancy between
the theoretical and experimental results. Firstly, the initial
condition of the condensate may be quite different from Gaussian due to
nonlinear and finite temperature effects. Condensates may begin with a small,
random center-of-mass momentum, giving an apparent dephasing effect
shot-to-shot. Alternatively, the condensate may have larger momentum width due to
heating or other effects.

We have estimated the momentum width necessary for an exponentially decaying fit
to the derived expression to return the same decay time as the experiment.
This yields $\sigma_k \approx 1.75 \times 10^6$ m$^{-1}$, which is 41\% greater
than our prediction. Approximately twice the predicted initial kinetic energy is
required to explain the experimental results.

\subsection{Nonlinear Regime}

\begin{figure}[t]
\begin{centering}
\includegraphics[width=7.5cm]{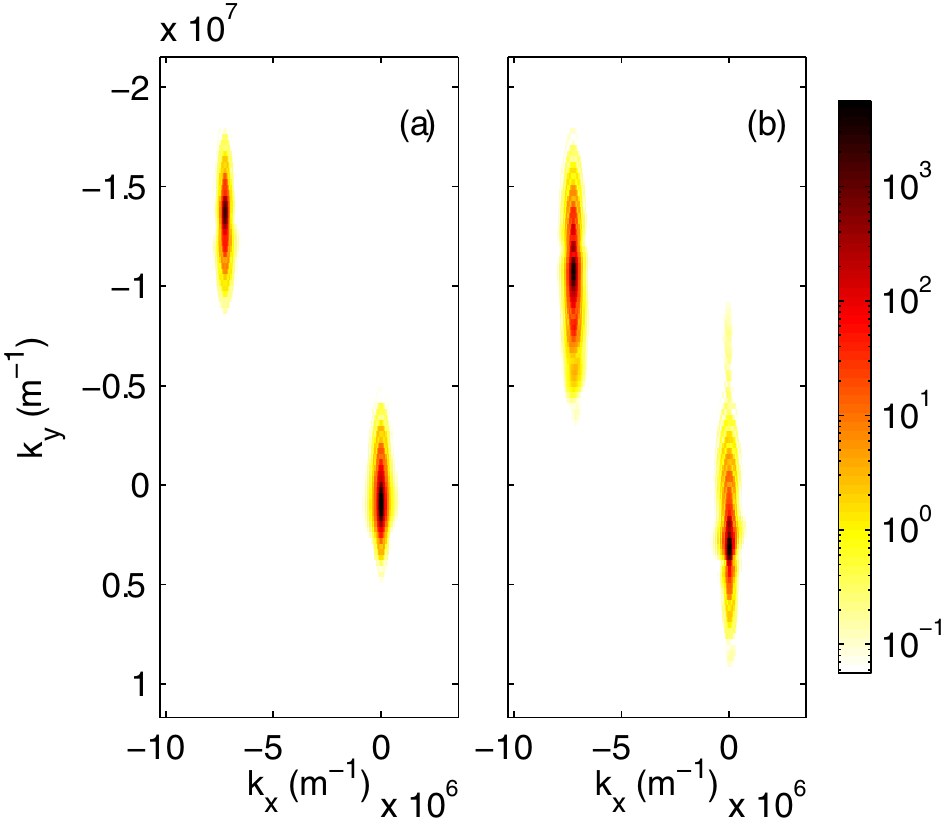}
\caption{Population per mode in momentum space is shown (a) 0.5 ms
and (b) 1 ms after the lattice is applied in a three-dimensional GPE
simulation. The $x$-axis is the weak trapping axis. The condensate
is visible at zero momentum and $2\hbar k_0$, but no thermal
features are visible. Note the ``camera'' angle is different to the
experiment and a logarithmic scale is used.} \label{gpe2d}
\end{centering}
\end{figure}

\begin{figure}[t]
\begin{centering}
\includegraphics[width=7.5cm]{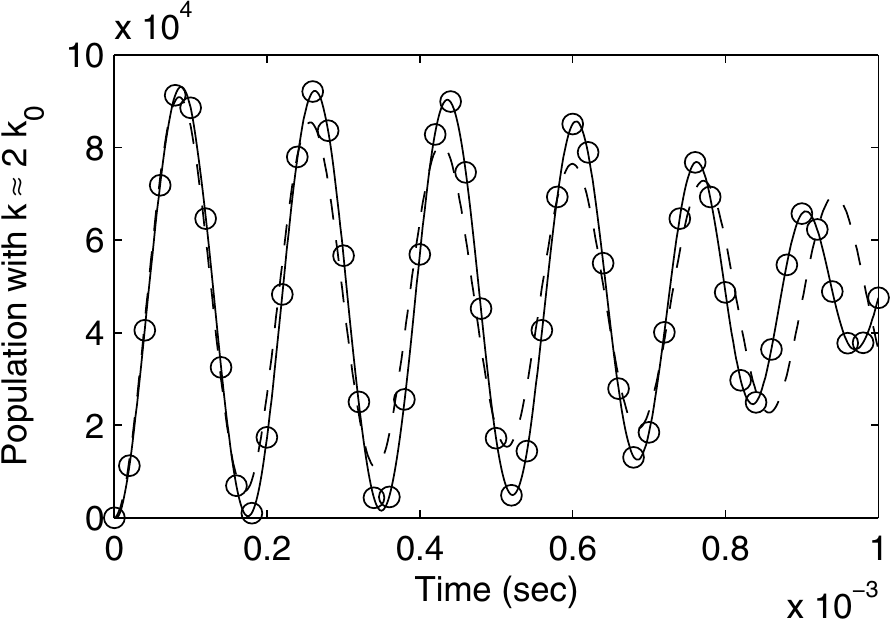}
\caption{Total population with positive momentum in the lattice
frame for a full 3D GPE calculation corresponding to the in trap
experiment showing damped Rabi oscillations. A cubic spline (solid
line, to guide the eye) and an exponential decay (dashed line) were
fitted to the simulation results (circles). The exponential decay
constant is $\tau = $1.3 ms.} \label{gperabi3d}
\end{centering}
\end{figure}

We now apply the Gross-Pitaevskii theory to simulate the
condensate dynamics in the combined magnetic and optical potential.
We perform a full 3D calculation directly matching the experimental parameters
described earlier. The initial condition for the calculation is generated via
imaginary time evolution to find the ground state containing approximately $1
\times 10^5$ atoms.

In Fig.~\ref{gpe2d} we display the momentum distribution after the
lattice applied to the trapped atoms for (a) 0.5 ms and (b) 1 ms
respectively . The broad, thermal features observed in the
experiment are not apparent in the simulation. However, small
ripples in the condensate momentum densities are indications of
linear dephasing. We observe that the momentum distribution of the
condensate modes broadens slowly in time. The center of mass of the
two condensates also evolve in time as the condensates climb the
walls of the trapping potential.

In Fig.~\ref{gperabi3d} we plot the populations with positive
velocity relative to the moving lattice. An exponentially decaying
sinusoidal fit results in a decay time of 1.3 ms. We note that the
exponential decay fits poorly to the numerical results. Qualitatively, some
features of the shape of the decay envelope matches more closely to
that derived in Eq.~(\ref{linearfitting}). In particular, the decay
is slower than exponential for short times and there is a notable
phase shift (or a change in frequency) reminiscent of the second
term inside the cosine of Eq.~(\ref{linearfitting}).

In the experiment we were able to separate the evolution of the
condensed atoms from the non-condensed, as shown in
Fig.~\ref{fig_exp_2} (b). The damping of the condensate oscillations
is at least partly caused by the linear dephasing. Linear dephasing
is present in this simulation, and should be comparable to that of
the experiment.  However, we see that the numerical result of 1.3 ms
is slower than the experimental fit of 800 $\mu$s. More favorable
agreement was found in Ref. \cite{Katz2004a} where the authors
applied a similar comparison with their experiment.

The Gross-Pitaevskii simulation did not generate the thermal
features found in the experiment. However, semiclassical theory can
explain the occurrence of thermalization by the existence of
dynamical instabilities, as shall be discussed below.

\subsection{Dynamical Instabilities}

The Gross-Pitaevskii equation is a non-linear equation that is known
to exhibit dynamical instabilities.

Stationary states of the GPE may, under certain circumstances, be
unstable points of equilibrium. That is, perturbations to the state
may grow exponentially in time. Eventually, for any realistic
initial condition the growing perturbations will alter the state
significantly.

The presence of dynamical instabilities can be detected by
linearizing the GPE about the stationary state, or equivalently by
solving the Bogoliubov-de Gennes equations. Quasi-particle modes
with imaginary eigenvalues are those that undergo exponential growth
and are dynamically unstable. Alternatively, modes can have negative
eigenvalues indicating energetic (Landau or thermodynamic)
instabilities.

There has been significant theoretical and experimental research
into the presence of dynamical instabilities in condensates in
optical lattices \cite{WuNiuBattle, Cataliotti2003a, Cristiani2004a,
Fallani2004a, DeSarlo2005a, Abdullaev2001a, Konotop2002a, Wu2003a,
Menotti2003a, Machholm2003a, Scott2003a, Machholm2004a, Zheng2004a,
Modugno2004a, Nesi2004a, Modugno2005a,Iigaya2006a,Polkovnikov2004a,
Ruostekoski2005a, Polkovnikov2005a, GeaBanacloche2006a}. What is
certain is that dynamical instabilities cause the thermalization of
the condensate when placed at the band edge of an optical lattice at
high enough densities.

We will not repeat previous work, but simply point out that
quantitatively accurate simulations of the regime in this experiment have not
been performed as far as we are aware. This regime involves a condensate at the
band-edge of a relatively weak 1D optical lattice, where transverse excitations
play an important role (as discussed in Sec. \ref{sec_dyn_inst}). The reason
for this is the lack of spontaneous processes in deterministic, semiclassical
simulations. Accurate quantum dynamics in other regimes has been performed in
\cite{Polkovnikov2004a,Ruostekoski2005a, Polkovnikov2005a}.

The exponential growth of quasi-particle modes is a result of Bose
enhanced (stimulated) processes. However, even with zero initial population
in the quasi-particle mode there is the chance of spontaneous
scattering into that mode. These spontaneous processes are very
important when determining the subsequent dynamics of the
condensate. In fact, in some regimes the spontaneous processes
dominate over the stimulated ones, as discussed in the next section.

\section{Regimes of dynamical instability in one-dimensional lattices}

\label{sec_dyn_inst}

In this section we discuss the effect of dynamical instability on
the condensate when one considers the spontaneous processes that may
occur.

The character of the dynamical instability in a lattice is dependent on many
factors, including the interaction strength, the number of particles, the
strength of the lattice and the dimensionality of the system. The role of
excitations in the directions transverse to the lattice was studied thoroughly
in Ref. \cite{Modugno2004a}. They showed that, for small enough $s$, that there
was a significant difference in the stability diagram for 1D and 3D models.

In our system, where $s \approx 3.1$, we would expect transverse
excitations to be important features of the dynamical instability.
Our eventual goal is to quantitatively describe the dynamics of the
condensate and so eventually must perform a full three-dimensional
analysis.  However, we will first discuss the dynamical instability
in systems of lower dimensionality.

For quasi-one-dimensional systems, one can expect a narrow band of
modes for which collisions are both resonant (i.e. conserve energy)
and conserve momentum modulo $2\hbar k_L$ (i.e. quasi-momentum in
the lattice). These are the modes which will undergo rapid
exponential growth. For the case of a pure condensate, the process
is initiated by spontaneous collisions between condensate particles. Then,
as the number of particles in the resonant modes grow,
Bose-enhancement of the scattering process causes parametric
amplification. The timescale for  this growth to become
significant will depend on the initial rate of spontaneous
scattering, and so should be included in any model which attempts
to quantitatively model dynamics. At later times, secondary
collisions between atoms will lead to thermalization of the system.

For  higher-dimensional systems there will be
a much larger volume in phase-space where collisions are close to
resonance, most of which incorporate an excitation transverse to the lattice
dimension. In this case the total rate of
condensate depletion by spontaneous scattering is greatly increased.
However, as there are many more modes the effect of Bose-enhancement
is significantly reduced. For some systems the number in each mode
may not grow much at all before the condensate is entirely depleted.
In this situation the parametric nature of the dynamical instability
will be washed out by spontaneous processes. Note that secondary
collisions would be expected in these systems too, which will lead
to thermalization. Thus, in systems of any dimensionality one could
expect to observe rapid thermalization of the condensate.

As an example, consider a system in the limit of a weak lattice
(small $s$). In this case the dispersion spectrum is not modified
significantly from the free particle case.  However, oscillations in
momentum due to the lattice are expected to occur. At the band-edge
one would expect Rabi oscillations between modes with momentum $\pm
\hbar k_L$. Assuming that the dispersion spectrum is not
significantly different to free space, one would expect resonant
collisions between the modes with momentum $\pm \hbar k_L$ into a
spherical shell of momentum with modulus $\hbar k_L$
\cite{Vogels2002a, Norrie2005a,Norrie2006a}. We can derive the
spontaneous collision rate using Fermi's second golden rule.

For a system with a total of $N \approx 10^5$
particles in a volume $V \approx 10^{-15}$ m$^3$
the spontaneous scattering rate is:
\begin{equation}
  \frac{d N}{dt} = \frac{U_0^2 N^2 m |k_L|}{2 \pi \hbar^3 V} \approx
  4\times10^7 s^{-1}, \label{fermi}
\end{equation}
where we have substituted values appropriate to our experiment. Thus, 10\% of
the condensate is lost in just 250~$\mu$s by spontaneous collisions.

Of course, this value will not be accurate for $s \approx 3$, and
does not take into account that the atomic populations are
oscillating between the two momentum states or the effect of
Bose-enhancement. This does however highlight that rapid
thermalization does not require parametric growth in dynamically
unstable 3D systems, but could be simply provided by a sufficiently large
density of states at resonance.

%\color{green}In an experiment one often creates two momentum components and free
%them from the trap. For some time we see spontaneous collisions - at
%later times the density decreases and the collision rate
%drops \cite{Norrie2005a,Norrie2006a}. This sort of calculation does
%not apply for a trapped condensate. In the Otago experiment, the atoms remain
%trapped. In that case the
%densities remain high and scattering continues for longer,
%and increasing via Bose-enhancement. \color{black}

\section{Quantum Model}

\label{sec_model}

The second-quantized Hamiltonian for an ultra-cold Bose gas is:
\begin{align}
  \hat{H} = & \int \mathrm{d}^3\mathbf{r}
  \left[ \hat{\psi}^{\dag}(\mathbf{r}) \left(-\frac{\hbar^2}{2m}
  \nabla^2 + V(\mathbf{r},t) \right) \hat{\psi}(\mathbf{r}) \right.\nonumber  \\
  &\left. + \frac{U_0}{2} \hat{\psi}^{\dag}(\mathbf{r})
  \hat{\psi}^{\dag}(\mathbf{r})  \hat{\psi}(\mathbf{r}) \hat{\psi}(\mathbf{r})
\right]  ,
  \label{fieldhamilitonian}
\end{align}
where the field operator $\hat{\psi}(\mathbf{r})$ annihilates an
atom of mass $m$ at position $\mathbf{r}$ and obeys the commutation
relation $[\hat{\psi}^{\dag}(\mathbf{r}), \hat{\psi}(\mathbf{r}')] =
\delta^3(\mathbf{r} -\mathbf{r}')$.

It is not possible to analytically solve  for the evolution of a Bose gas
system directly from the Hamiltonian in
Eq.~(\ref{fieldhamilitonian}). Analytical approaches always involve
approximations, and often linearize the quantum
fluctuations about the mean-field as in the Bogoliubov
approach. Numerical solutions are possible with the exact Positive-P
method, but quickly become intractable with sampling problems in many situations
\cite{steele1998}. The approximate Gross-Pitaevskii equation (GPE)
is straightforward to solve, but fails to account for any
spontaneous effects that are important for the problem we are
considering.

In this paper we implement the truncated Wigner phase-space method
for the system dynamics. This is an approximate method that
goes beyond the GPE by including both spontaneous processes and the
dynamics of non-condensate modes.

\subsection{The Truncated Wigner Method}

\label{sec_tw_method}

We will now briefly explain the origin and use of the truncated Wigner method.
This was used in the context of squeezing of solitons in optical fibres by
Carter and Drummond \cite{Carter1987a}, and was first applied to Bose gases by
Steel \emph{et al.} \cite{steele1998}.  An equation of motion for the Wigner
function can be obtained from the Hamiltonian Eq.~(\ref{fieldhamilitonian}) by
deriving the master equation for the system followed by the appropriate
transformation of the density operator. However, for a system with $n$ modes the
Wigner function is $2n$-dimensional and it is impractical to solve such an
equation directly.

However, truncating the third-order derivatives in the equation of motion for
the Wigner function results in a Fokker-Planck equation, which can then be
simulated using stochastic methods \cite{Gardiner1985a,gardiner1999a}. The
omission of the higher order terms  has previously
been justified
when the there are more particles than modes in a calculation and the simulated
times are short~\cite{Sinatra2002a,Norrie2005a,Norrie2006a}. The equation of motion
for the stochastic trajectories is simply the usual Gross-Pitaevskii equation,
although the initial conditions for the field are sampled stochastically from
the appropriate Wigner distribution.  A large number of trajectories must be
calculated, which are then sampled to
obtained expectation values of symmetrically-ordered quantum field operators.

As an example, the ensemble average $\overline{\psi^{\ast} \psi}$ is
equal to the expectation of the symmetric ordering of the number
operator, or $\langle\hd{\psi}\h{\psi} + \h{\psi}\hd{\psi}\rangle/2$. Therefore,
the average number of particles in a mode is $\langle \hd{\psi}
\h{\psi} \rangle = \overline{\psi^{\ast} \psi} - 1/2$.  For more
details on the truncated Wigner and other phase space methods see
\cite{steele1998,Gardiner2003,gardiner1999a}.

Because of the nonzero width of the Wigner distribution, every vacuum
mode begins with uncorrelated Gaussian noise, normalized to an
average of half a particle per mode. This provides the seeding
needed for the equation of motion (the GPE) to allow scattering
events into these modes. Furthermore, the rate of scattering
corresponds to that of the expected spontaneous processes.

Allowing such spontaneous processes is essential to accurately model
systems exhibiting instability. Unstable dynamical solutions of the
GPE do not correspond accurately to quantum mechanics. Such states
would eventually be altered by spontaneous processes and are
therefore not equilibrium states.
The truncated Wigner approach unambiguously allows both dynamical
instabilities and Landau instabilities to manifest.
Additionally, a background thermal gas can be included in the
initial condition to allow dissipation via the Landau instability.
The importance of the two instabilities has been a point of
discussion in the past \cite{WuNiuBattle}.

The disadvantage of the truncated Wigner method is the accuracy of
its fundamental approximation is difficult to quantify. However,
some work has been done which demonstrates its validity regimes
\cite{Sinatra2002a,Norrie2006a,Deuar2006a,Polkovnikov2003a,Polkovnikov2007a}.
Some of these works have concluded that the major requirement
for accuracy is that the number of modes must be much smaller than the number of real
particles \cite{Sinatra2002a,Deuar2006a}. If this is not the case,
then the ``vacuum noise'' in the initial condition can no longer be
considered a perturbation. The vacuum fluctuations will begin to
interact as if they are real particles, and unrealistic effects such
as negative expectation value for population become prolific
\cite{Deuar2006a}.

The method in \cite{Polkovnikov2003a,Polkovnikov2007a} allows one to
check the validity of the truncated Wigner approximation at the
expense of significant extra computation. In this paper, however,
we instead choose
to work in the regime where there are more particles than simulated
modes.

Unfortunately, this is not the case for the full 3D experimental parameters.
Including the trap and the optical lattice whilst simultaneously avoiding
numerical aliasing due to the nonlinearity requires of order $10^6$ modes to
represent  $1 \times 10^5$ atoms. Our calculations in this regime clearly
display unphysical behavior similar to those found in \cite{Deuar2006a}. In
particular, we observe large areas of negative population at high momentum
modes, and a corresponding clustering of the vacuum noise at small momentum
which swamps the condensate. The condensate is rapidly depleted as it scatters
off these unphysical particles.

To proceed we must make a further simplification by neglecting the
trapping potential, and instead  modeling a homogeneous system with a
similar density to the experiment. By doing so we expect to capture
the most important physics, as the timescale that the thermalization
manifests is much shorter than the inverse trapping frequencies, so
that whilst the optical lattice is on the atoms will be essentially
stationary and the density constant. The most important effect of
the trap in the experiment is to maintain the high density so that
collisions may occur.  This model requires fewer modes to simulate
the dynamics, and therefore the truncated Wigner method is expected
to be more accurate.

In the next section we describe the details and present the results from our
truncated Wigner simulations. We point out in advance that while these
calculations will model the spontaneous scattering in this system,  the
homogeneous calculations will not capture the linear dephasing of the Rabi
oscillations in the experiment as the condensate begins in a
single momentum state. We will consider this further in Sec.~\ref{sec_thermal}.

\section{Truncated Wigner simulations}

\label{sec_tw_results}

In the experiment the initial inhomogeneous condensate wave function spreads
over about 27 wells of optical lattice potential.
We use a total population of $10^5$ atoms and choose to use a density of
$n_0 = 10^{20}$ m$^{-3}$, roughly corresponding the the average density within
the trap rather than the peak density. We have implemented a
rectangular grid with periodic boundary conditions for
our simulation space, with dimensions $L_x \times L_y \times L_z$.
The lattice is parallel to the long axis of the space, with spacing 390 nm,
extending over 32 wells, making $L_x = 12.48$ $\mu$m.
It follows that $L_y = L_z \approx 8.95$ $\mu$m. Note that this
geometry is slightly different to that of the extended
condensate at a $63^{\circ}$ angle to the lattice.

We begin with a homogeneous cloud moving at the Bragg condition with
momentum of $\hbar k_L$ in the direction of the stationary lattice.
The nonlinearity and strength of the lattice are the same as used
previously. Care was taken to ensure numerical accuracy, and
a projection operator that prevents numerical aliasing
\cite{Davis2001a, Davis2001b, Davis2002a} was employed. %Comparisons
%indicate that the grid size used was sufficient to make aliasing
%effects negligible.

In the following subsections we present our results of the truncated
Wigner method used to simulate the experiment using 1D, 2D and 3D
models. This was done for two reasons. The first was to demonstrate
the different regimes discussed in Sec.~\ref{sec_dyn_inst}.
Secondly, the truncated Wigner method will be very reliable in the
1D and 2D models where the number of modes is much smaller than the
number of particles.

To compare the simulations at different densities we have ensured
that the nonlinearity has scaled correctly. In 1D, the nonlinearity
is $U_0/(L_y L_z)$ and in 2D it is $U_0/L_z$.

The one-dimensional results in Sec. \ref{sec_1d} clearly show the
dynamical instability and the parametric gain of the resonant modes.
The two-dimensional simulations in Sec.~\ref{sec_experimentd} allow
us to clearly visualize the transverse excitations, as was achieved
previously in \cite{Katz2005a}. We also analyze the coherence and
statistics of the field in 2D where statistical errors are more
manageable that the 3D situation, as more trajectories can be
computed for the same amount of CPU time. Finally, in
Sec.~\ref{sec_dyn_instd} we show the results of the complete
three-dimensional simulation that exhibits strong spontaneous
scattering and make a comparison to the experimental results.

\subsection{One Dimension}

\label{sec_1d}

\begin{figure}[t]
\begin{centering}
\includegraphics[width=8.05cm]{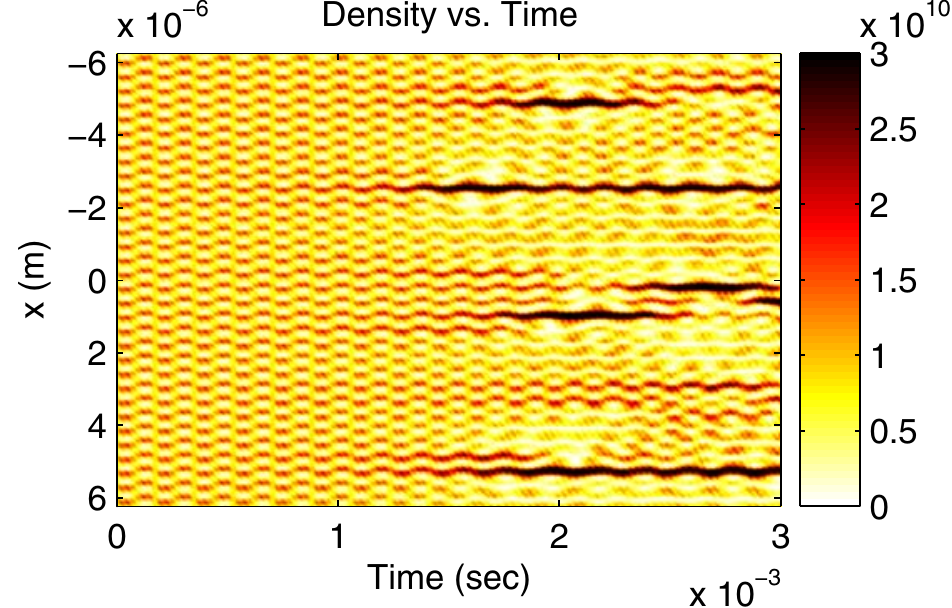}
\includegraphics[width=8cm]{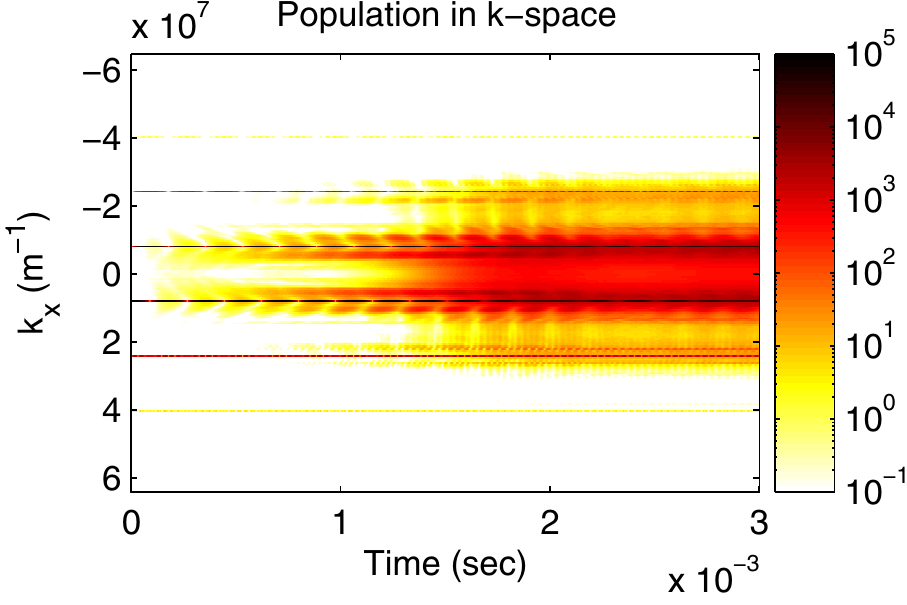}
\caption{(Color online) Top: Density of a 1D BEC moving in an optical lattice.
 We see spontaneous symmetry breaking of the BEC in
in a single trajectory simulation. The localized
  regions of high density are bright-gap solitons. Bottom: The same results
  are presented in momentum space, where symmetry breaking is
  characterized by mixing into a variety of different modes (note the
  logarithmic dependence of the density scale).}
\label{single1d}
\end{centering}
\end{figure}

\begin{figure}[t]
\begin{centering}
\includegraphics[width=7cm]{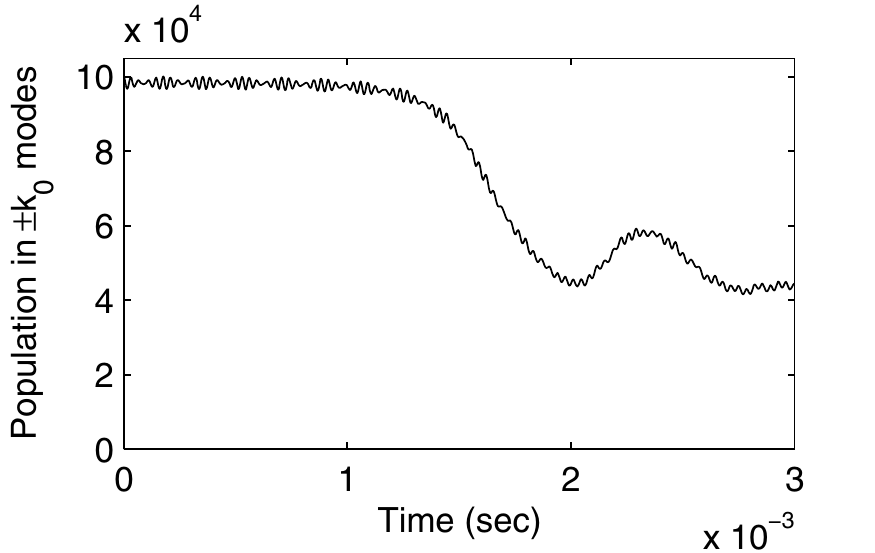}
\includegraphics[width=7cm]{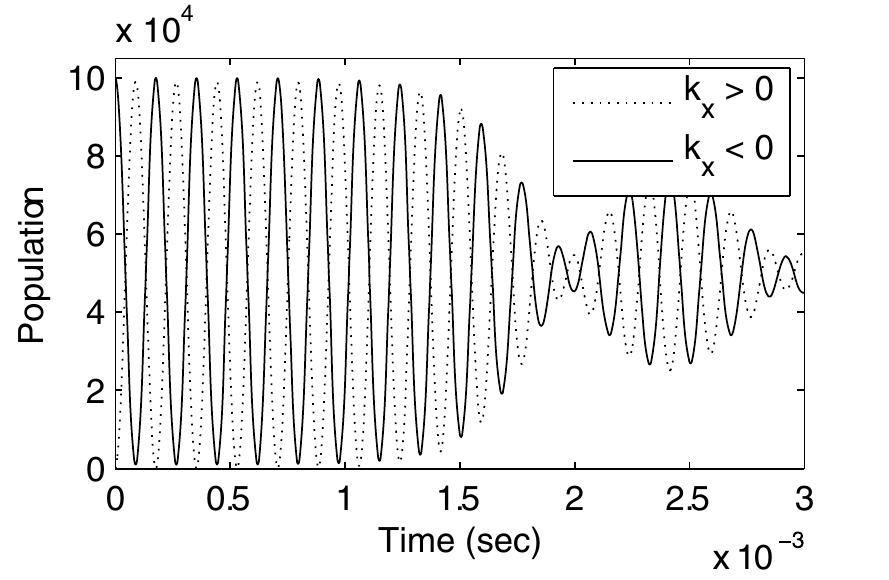}
\caption{Top: Sum of the population in the momentum modes $\pm k_L$ of a
  1D Bose gas
  plotted versus time. The small, rapid oscillations are due to the
  initial state mixing with higher band states in the Bloch
  basis. The atoms are depleted from the condensate mode, until a
  partial revival begins at $\sim 2$ms. This is due to Rabi oscillations of
  the not-quite resonant collisions. Bottom: Total
  populations of atoms with positive or negative momentum versus time. Damping
  of the Rabi oscillations is
  observed in the 1D simulations after 1.5ms, but then revive at the same time
  as the condensate as shown in the top figure. Both plots have used an
  ensemble of 1000 trajectories.}
\label{interference}
\end{centering}
\end{figure}

We first investigate the dynamical instability in one-dimension.
Figure \ref{single1d} shows results for a single trajectory, which in
some sense can be considered to be analogous to what might be
observed in a single experiment in the lab \cite{Norrie2006a}. In
real space, we observe spontaneous breaking of the translational
symmetry of the lattice brought about by the dynamical instability.
The high-density features correspond to generation of bright-gap
solitons. These solitons are known to be generated from states near
the top of the ground band for condensates with repulsive
interactions \cite{Konotop2002a}. The momentum distribution broadens
indicating spontaneous collisions between condensate atoms
scattering atoms to other momentum modes.

The work of Hilligs{\o}e and M{\o}lmer \cite{Hilligsoe2005a,Molmer2006a}
predicts that collisions that conserve energy and momentum (i.e.
resonant collisions \textit{in the eigenbasis of the lattice}) will
experience parametric growth. If energy conservation is not quite
satisfied for a particular mode then observe Rabi oscillations are expected
that will hinder
the exponential growth of parametric amplification. This is
exactly what we observe in the simulation as the resonance condition is
never precisely matched. This effect is demonstrated in
Fig.~\ref{interference}, where population revival into the states
${\pm}k_L$ can be seen. Also note that only about half the
population is depleted from the two condensate modes -- this corresponds
to the half that are in the lowest band near the band-edge that
generate a bright-gap soliton \cite{Konotop2002a}. %\mjd{(???? MJD)}

The experiments described in this paper indicate damping of the Rabi
oscillations is due to dynamical instability. Our 1D simulations
show damping occurring after a few milliseconds (see
Fig.~\ref{interference}). Ensemble averages of 1000 trajectories
have shown the spread of momentum states and the approach towards
thermalization. It should be noted that phase coherence is also
destroyed by the dynamical instability. This is an interesting
example of decoherence under Hamiltonian evolution.

% I wonder whether its worth plotting |g^(1) (r,r + dr)| vs dr?

\subsection{Two Dimensions}

\label{sec_experimentd}

\subsubsection{Dynamical instability}

\begin{figure}[t]
\begin{centering}
\includegraphics[width=7.5cm]{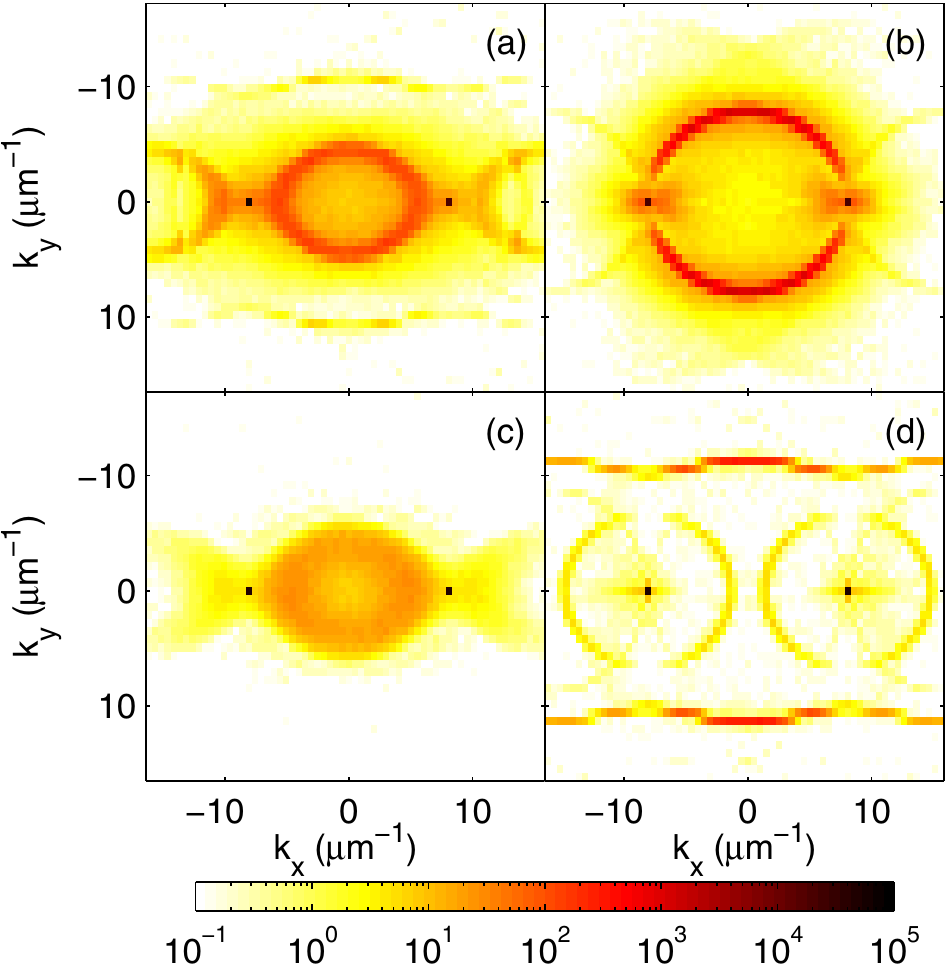}
$\left\langle\hd{a}\h{a}\right\rangle(\mathbf{k},t)$
\caption{(Color online) Populations in
  momentum space of a 2D simulation. (a) The condensate began in the
  lattice in the $-k_L$ momentum state, $t$~=~1.1~ms. (b) The condensate
  began in a superposition of $\pm k_L$ momentum in free-space for
  comparison with (a), $t$~=~1.1~ms. (c) The condensate began in the lowest
  band-edge of the lattice, $t$~=~0.36~ms. (d)  The condensate began
  in the edge of the first excited band, $t$~=~1.1~ms. Movies of these
  simulations are available online at:
  http://www.physics.uq.edu.au/people/ferris/.}
\label{plots2d}
\end{centering}
\end{figure}

\begin{figure}[t]
\begin{centering}
\includegraphics[height=3.15cm]{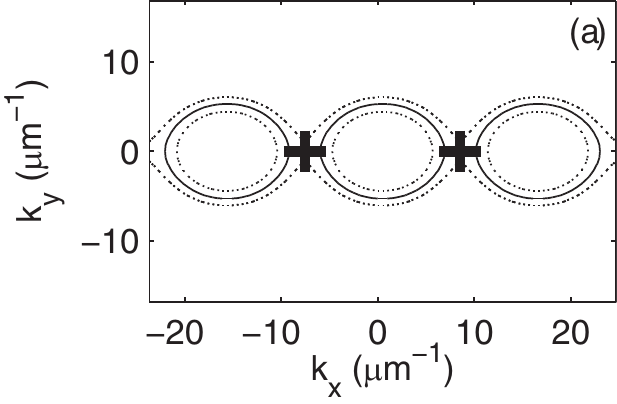}\hspace{-0.06cm}%
\includegraphics[height=3.15cm]{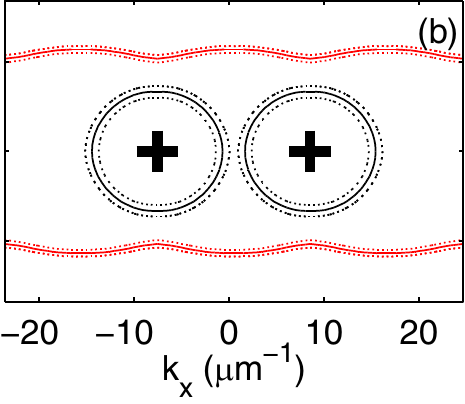}
\caption{(Color online) Plots of modes for which the resonance condition is
  satisfied. The dotted lines correspond to regions close to
  resonance, having an energy mismatch corresponding to a Rabi
  oscillation period of 1~ms. The large pluses represent the initial state of
  the BEC. (a) Two atoms colliding within the lowest band. (b)
  Two atoms begin in the second band-edge can collide to result in an
  atom in the lowest and second band (inner circles, black online) or
  two atoms in the lowest band (outer rings, red online).}
\label{resonance}
\end{centering}
\end{figure}

As described above, the effect of dimensionality on the system varies
according to the parameters of the experiment, and has been dealt
with analytically in detail in \cite{Modugno2004a}. In that
treatment, the dynamical instability was studied as an exponential
growth of Bogoliubov modes. A complete set of Bogoliubov modes
includes excitations in the direction of the lattice (such as would
be found in a 1D treatment), excitations in the directions
perpendicular to the lattice, and excitations which are mixtures of
both. For some parameter regimes, the perpendicular excitations are
an important part of the dynamics.

The same is true in our homogeneous treatment. Collisions may be
resonant into modes with non-zero components of momentum in the
directions perpendicular to the lattice. This can be observed quite
clearly in the results of our two dimensional simulations.

Figure \ref{plots2d} shows a comparison momentum distributions
for four different initial conditions averaged over $10^3$ trajectories. In
Fig.~\ref{plots2d}(a) the condensate begins in the $+k_L$ momentum
mode; in Fig.~\ref{plots2d}(b) we make a comparison to free space
collisions between momentum modes $\pm k_L$. Figs.~\ref{plots2d}(c)
and (d) begin with the atoms loaded into the ground and first
excited band-edge states.

When the moving lattice is switched on such that the BEC is in the
$+k_L$ momentum state, we see complicated dynamics including Rabi
oscillations and scattering into a variety of modes. A significant
percentage of the atoms have developed momentum in the direction
perpendicular to the lattice, so we obviously can not characterize
the system as quasi-1D.

We can understand the complicated dynamics by realizing that the
initial condition is a superposition of momentum states as given in
Eq.~(\ref{BlochStates}). Each of these states has a well-defined
energy, and we can analyze which collisions will conserve energy and
quasi-momentum (modulo $2k_L$) in the Bloch basis. The system has
been analyzed previously in a similar fashion
\cite{Katz2005a,Molmer2006a}. Katz~\textit{et al}. \cite{Katz2005a}
provided comparisons with the results of truncated Wigner
simulations and an experiment. They found, as we do, that the
structure of the scattering halo differs significantly from the
s-wave scattering sphere generated by condensate collisions in free
space, due to the modified dispersion curve.

Two atoms initially at the edge of the first band can collide to
produce atoms elsewhere in that Bloch band, but with additional
kinetic energy in the transverse direction. We have calculated the
resonance condition using the Hartree-Fock mean field method
\cite{PethickSmith}. The values of momentum where this occurs is
shown in Fig.~\ref{resonance}(a) (\textit{cf}.
Fig.~\ref{plots2d}(c)). x There are two possible outcomes for atoms
colliding at the first excited band edge. Either one or both atoms
can end up in the lowest band. Both of these cases are presented in
Fig.~\ref{resonance}(b) [\textit{cf}. Fig.~\ref{plots2d}(d)]. Note
that the resonance condition is sharper than in Fig
\ref{resonance}(a), i.e. there is a smaller area of modes which have
a final energy close to the resonance condition. This explains the
results in Fig.~\ref{plots2d}, where the lowest band-edge initial
condition exhibits significantly more scattering at short times than
the second band-edge case.

\subsubsection{Coherent and incoherent components}

\begin{figure}[t]
\begin{centering}
$\hspace{1cm} N_{\mathrm{coherent}}(\mathbf{k},t) \hspace{1cm}
N_{\mathrm{incoherent}}(\mathbf{k},t)$
\includegraphics[width=7.5cm]{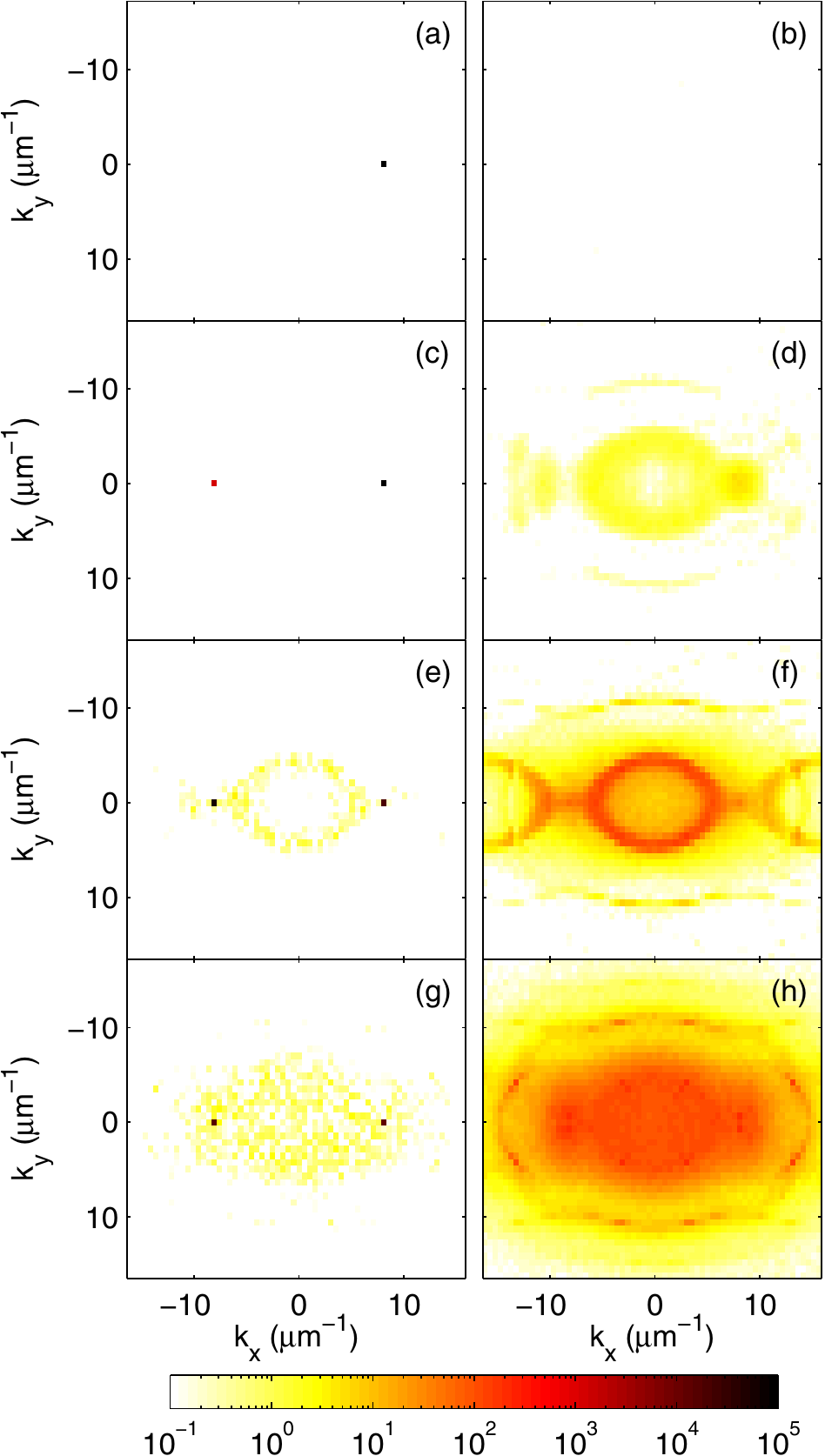}
\caption{(Color online) Coherent (a,c,e,g) and incoherent (b,d,f,h) populations
in momentum space for the 2D simulation at times 0 (a,b), 0.56~ms (c,d), 1.1~ms
(e,f), and 3~ms (g,h).}
\label{fig:coherent2d}
\end{centering}
\end{figure}

We have investigated the phase coherence of the atomic cloud as it
undergoes thermalization. A Bose-Einstein condensate has
off-diagonal, long-range order where there is a well-defined
relative phase across the system. Any overall phase of the
condensate results from spontaneous symmetry breaking, and the value
of the global phase is undetermined.

Nevertheless, when employing classical field methods one typically begins with a
coherent state, ascribing a global phase to the condensate at $t=0$ such that:
\begin{equation}
  \left|\left\langle \h{a}_0 \right\rangle\right| ^2 = \left\langle
  \hd{a}_0 \h{a}_0 \right\rangle ,
\end{equation}
where $\h{a}_0$ is the annihilation operator for the condensate
mode. All observables of the closed
system being simulated will return the same results if one averages over
ensembles of random global phase, that is averaging $\arg \langle \h{a}_0
\rangle$ from 0 to $2\pi$.

The value of the phase of $\langle\h{a}_0\rangle$ is only meaningful
in respect to a second condensate of atoms. An interference
experiment can resolve the phase relationship of two condensates. If
one condensate undergoes heating and becomes thermalized in an
experiment, then that cloud of atoms will lose its coherence and not
result in fringes in an interference experiment. For a thermal
sample $\langle \h{a} \rangle = 0$.

We define the \textit{coherent} component of the cloud to be that
with well-defined global phase
\begin{equation}
  N_{\mathrm{coherent}}(\mathbf{k}) = \left|\left\langle
  \h{a}(\mathbf{k}) \right\rangle\right|^2  .
\end{equation}
The incoherent component is the remainder of the atoms
\begin{equation}
  N_{\mathrm{incoherent}}(\mathbf{k}) = \left\langle
  \hd{a}(\mathbf{k})\h{a}(\mathbf{k})\right\rangle -
N_{\mathrm{coherent}}(\mathbf{k}).
\end{equation}

We plot the coherent and incoherent populations in
Fig.~\ref{fig:coherent2d} for the initial condition with the
condensate in  momentum state $|\!+\!k_L\rangle$. Coherent
dynamics caused by the lattice drives the condensate into the modes
with momentum $\pm k_L$, $\pm 3 k_L$, etc. Only these modes contain
any significant coherent population. Note the noise in
Fig.~\ref{fig:coherent2d} (e,g) is a statistical artefact from averaging over
only 1000 trajectories, and could be reduced with more CPU time.

The incoherent population is generated by spontaneous scattering into
the non-condensate modes surrounding the condensate. At short times
energy conserving collision from the two condensate modes dominates. At
later times further scattering from the newly populated modes broadens
the distribution and results in thermalization.

\subsubsection{Local correlation functions}

\begin{figure}[t]
\begin{centering}
\includegraphics[width=7.5cm]{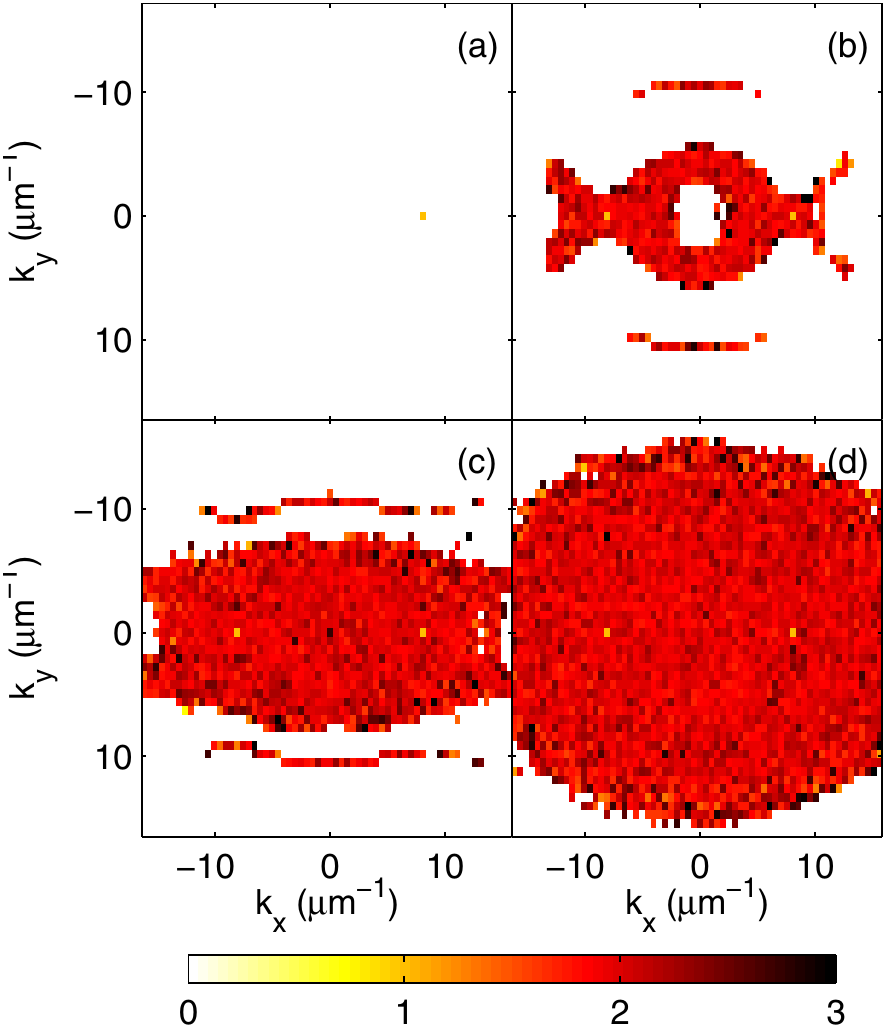}
$g^{(2)}(\mathbf{k},t)$ \caption{(Color online) The normalized
second order momentum-space correlation function
  $g^{(2)}(\mathbf{k})$ is shown for the 2D simulation at times (a) 0 ms,
  (b) 0.56 ms, (c) 1.1 ms, and
  (d) 3.0 ms. Results are only shown for modes with average
  population greater than 1/2.}
\label{fig:g2}
\end{centering}
\end{figure}

The normalized second-order local momentum-space correlation
function is:
\begin{equation}
  g^{(2)}(\mathbf{k}) = \frac{\left\langle \hd{a}(\mathbf{k})
  \hd{a}(\mathbf{k}) \h{a}(\mathbf{k}) \h{a}(\mathbf{k})
  \right\rangle}{\left\langle \hd{a}(\mathbf{k}) \h{a}(\mathbf{k})
\right\rangle^2},
  \label{eq:g2}
\end{equation}
and can be calculated using the truncated Wigner method by
transforming into symmetric ordering. This correlation function
allows us to probe the quantum statistics of each momentum mode --
in particular, the occupation statistics in each mode. For coherent,
Poissonian statistics $g^{(2)} = 1$, and for thermal, Gaussian
statistics $g^{(2)} = 2$.

We plot $g^{(2)}(\mathbf{k},t)$ in Fig.~\ref{fig:g2} for the state
beginning with momentum $k_L$. At all times $g^{(2)}(\pm k_L)
\approx 1$, and the condensate modes continue to display coherent
statistics. However, we see that the all other populated modes have
$g^{(2)} \approx 2$. These modes are displaying Gaussian statistics
which supports the claim that a thermal cloud is growing around the
condensate. Comparing Figs.~\ref{fig:coherent2d} and \ref{fig:g2} we
see the modes with phase coherence display second-order coherent
statistics and those without phase coherence display thermal
statistics. Note again that the noise in Fig.~\ref{fig:g2} is
statistical.

\subsubsection{Rabi oscillations}

\begin{figure}[t]
\begin{centering}
\includegraphics[width=7cm]{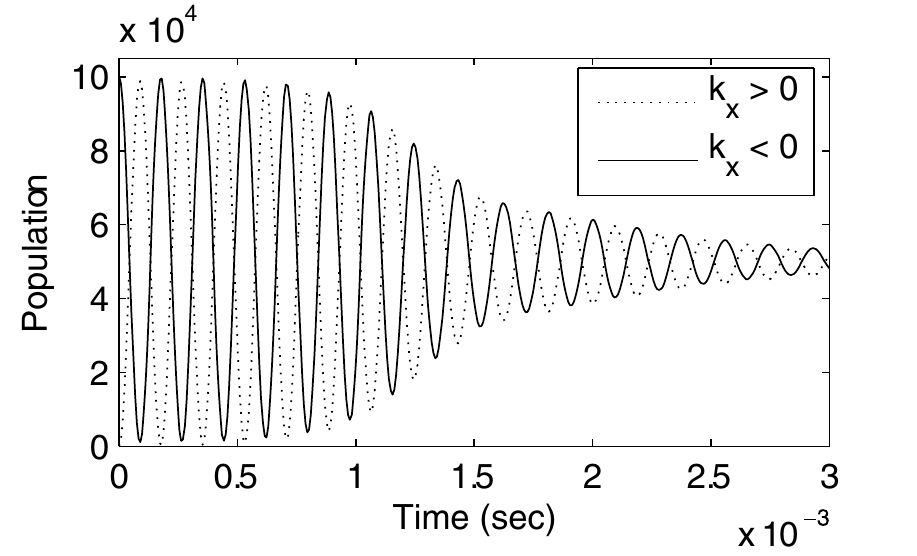}
\caption{Time dependence of the total population of all modes with positive
or negative momentum in the lattice direction (in the frame of the lattice).
These exhibit damped Rabi oscillations in the 2D simulations similar to
the 1D results in Fig.~\ref{interference}. Note that onset of
damping is more rapid in 2D than 1D, and that revivals are not seen
in this case. } \label{pops2d}
\end{centering}
\end{figure}

Finally in Fig.~\ref{pops2d} we present results the time dependence of the
positive and negative momentum components for the 2D system.  These show damped
Rabi oscillations with no revivals as observed in in the 1D simulations. There
are many more modes in the 2D system, which has two important effects. First,
each mode will Rabi oscillate at a different frequency, and so one expects
dephasing to wash out the revival. Second, these modes will interact with each
other, and the system will approach thermal equilibrium faster. We also see that
these the oscillations undergo faster damping compared to the 1D results.

\subsection{Three Dimensions}

\label{sec_dyn_instd}

\subsubsection{Numerical accuracy}

\begin{figure}[t]
\begin{centering}
\includegraphics[width=7cm]{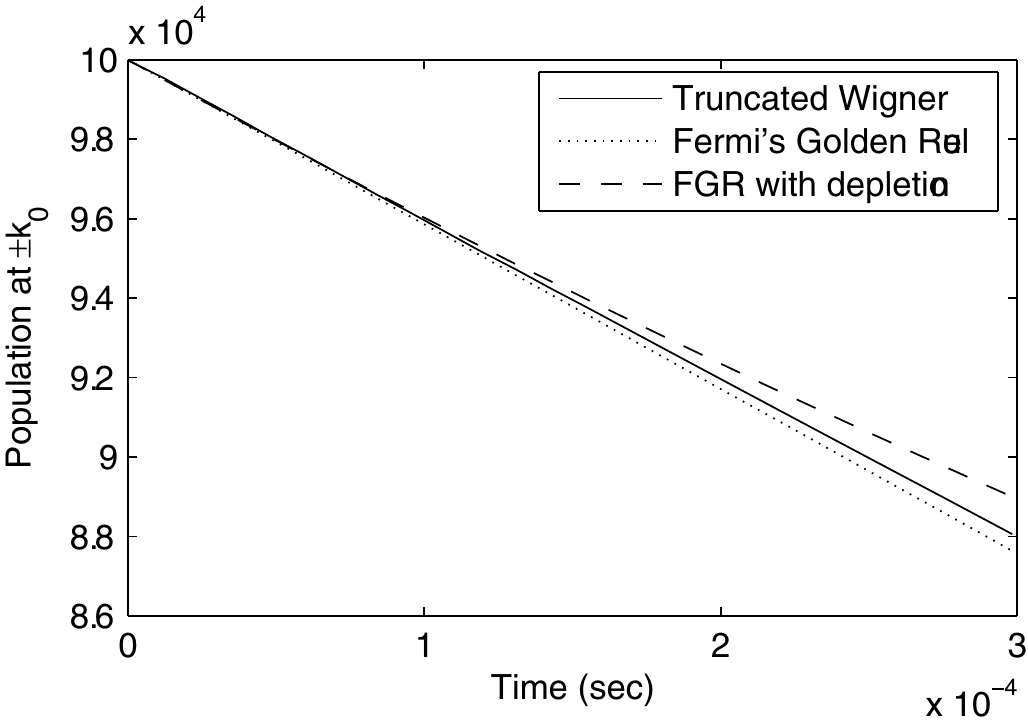}
\caption{Condensate occupation of the modes $\pm \hbar k_L$ versus
time for a condensate collision in free space.  We compare
predictions from truncated Wigner simulations (solid), a straight
line fit to the rate given by Fermi's golden rule (dotted), and the
solution to the differential equation Eq.~(\ref{fermi}) that
includes condensate depletion (dashed).} \label{depletion}
\end{centering}
\end{figure}

The truncated Wigner method for quantum dynamics is an approximation
that is only known to be accurate in certain regimes
\cite{Sinatra2002a,Norrie2006a,Deuar2006a,Polkovnikov2003a}. It is known to be
accurate under the condition that there are many more particles
in the simulation than modes.  For the experiment we are considering this
is difficult to achieve, as in
three dimensions the number of modes can easily be greater
than one million.
Thus, in our calculations we took great care to use the minimum
possible number of modes
without misrepresenting the physics. To this end we incorporated a
projector to implement a spherical
cut-off in momentum space at a radius slightly
greater than where the atoms can be expected to be found as
estimated from lower-dimensional simulations. A sufficiently large grid was used
in real space to eliminate the effects of numerical aliasing
\cite{Davis2001a, Davis2001b, Davis2002a}.
This reduces the number of modes to $\sim 4 \times 10^5$, for the
simulation of $10^5$ atoms. The truncated Wigner method will not be
accurate for long time scales under these conditions, but we are
confident that the dynamics we see in relatively short time scales
here is accurate.

We have not implemented the method presented by Polkovnikov
\cite{Polkovnikov2003a, Polkovnikov2007a} to check the validity of
our truncated Wigner simulation. Instead, we have tested our
accuracy by comparing our simulation of a simple situation with
known results. We can compare the spontaneous scattering rates from
a simulation with atoms with momentum $\pm \hbar k_L$ with Fermi's
second golden rule, Eq.~(\ref{fermi}). We have performed this
simulation for the same parameters, but without the lattice. The
number of atoms remaining in the $\pm \hbar k_L$ momentum modes is
plotted as a function of time
 in Fig.~\ref{depletion}. At
small times the results agree well. We can see the effect of
Bose-enhancement at later times when Fermi's golden rule
underestimates the depletion rate.

\subsubsection{Momentum distribution}

\begin{figure}[t]
\begin{centering}
\includegraphics[width=7cm]{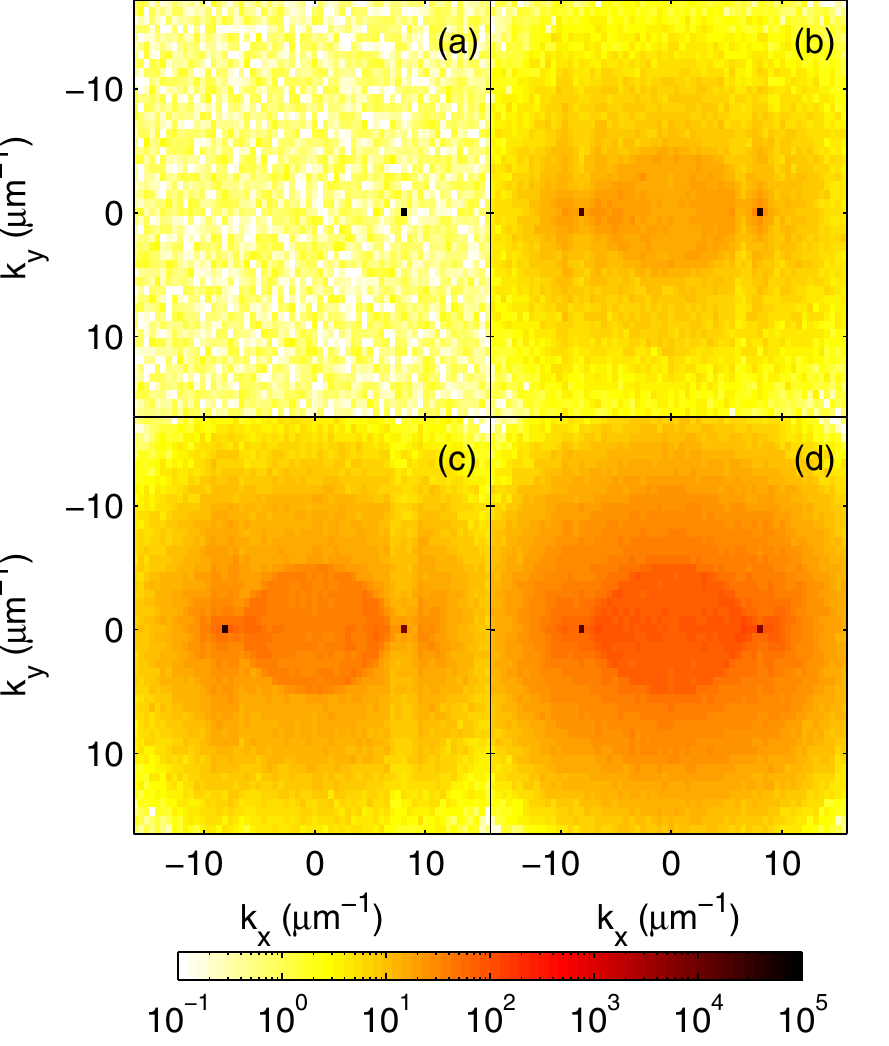}
\caption{(Color online) The momentum space column densities for the
3D simulation  is shown at times  (a) 0 ms,
  (b) 0.56 ms, (c) 1.1 ms, and
  (d) 3.0 ms.  The scattering halo is clearly visible in (c) and (d).} \label{plots3d}
\end{centering}
\end{figure}

\begin{figure}[t]
\begin{centering}
\includegraphics[width=8cm]{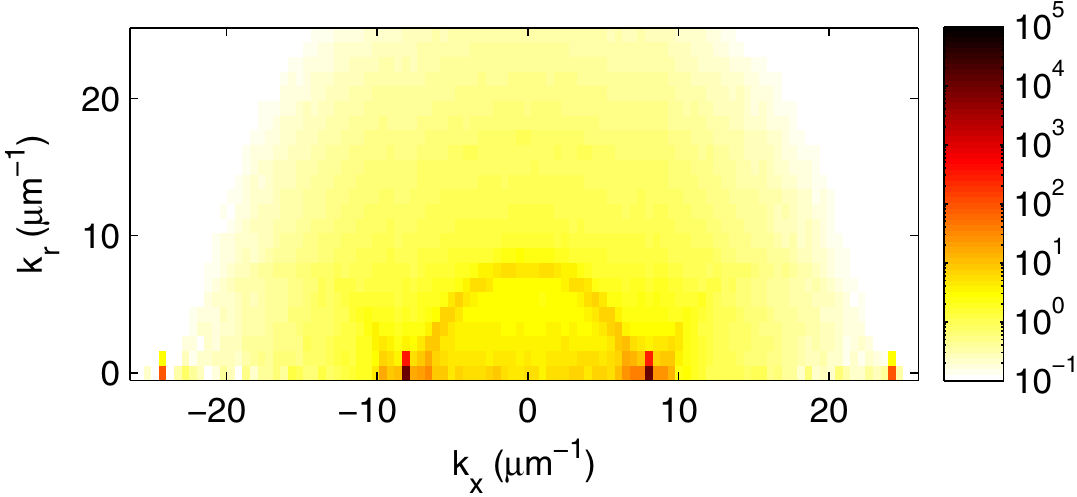}
\caption{(Color online) The average population per mode is shown as a function
of
  axial and radial momenta, for a 3D simulation at time 3.0 ms. }
\label{radial3d}
\end{centering}
\end{figure}

In the experiment, time-of-flight expansion was used to image the
momentum distributions of the condensates, giving 2D column densities.
Fig.~\ref{plots3d} shows the results
of the three-dimensional calculation where we have summed the
populations over the $z$ momentum component. The
ensemble average is for only 32 trajectories due to the computational
demands of the simulation. Note that the sharp features seen in the
2D case (Fig. \ref{plots2d}) are somewhat washed out, resulting in a
distribution more closely resembling two (compact) condensates
surrounded by a thermal cloud. However, this is not precisely the
case -- the population is slightly concentrated in ring-shaped structures in
3D as seen in Fig.~\ref{radial3d}. These ring-shape structures are
similar to those seen in the 2D simulations.

We remind the reader that the simulations performed here are for a homogeneous
gas,  whereas the condensates in the experiment has a finite
momentum width.  It seems reasonable to expect that the sharper features would
not be as visible in a trapped condensate. In experiments, the
expansion is done in finite time which can cause significant
broadening of the condensate peaks and other features. Thus, it is
reasonable to say there is good qualitative agreement between
Fig.~\ref{plots3d} and the experimental images in
Fig.~\ref{fig_exp_photo} when one takes the effects of the trapping
potential into account.

\subsubsection{Rabi oscillations}

\begin{figure}[t]
\begin{centering}
\includegraphics[width=7cm]{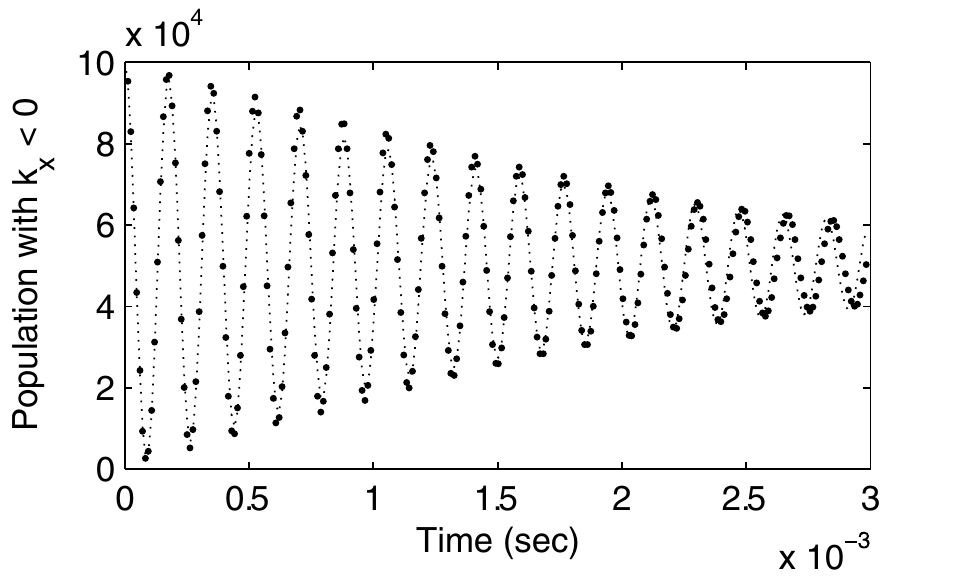}
\caption{Population of the momentum modes with $k_x <0$ for the 3D
simulations. We see damped Rabi oscillations similar to
Figs.~\ref{interference} and \ref{pops2d}. A decaying exponential
fit (dotted line) has been made with the numerical results (points),
with period 177 $\mu$s and decay time constant 2.00 ms. }
\label{pops3d}
\end{centering}
\end{figure}

We have plotted the total population with a positive component of
momentum in the lattice direction in Fig.~\ref{pops3d}. An
exponentially decaying oscillation of the form
\begin{equation}
  P_{+} = P_0 \cos(\omega t) e^{-t/\tau}
\end{equation}
was fitted to the data using a least-squares method. The oscillation
period was found to be $2 \pi/ \omega = 177$ $\mu$s, as expected for a lattice
with $s = 3.1$.

The decay constant was found to be $\tau = 2.00$ ms. This decay is
significantly slower than in the experiment, where the decaying fit
yields $\tau = 450$ $\mu$s. There are a number of important
experimental considerations that must be accounted for in addition to this
result. We shall discuss how to
estimate the effects of finite temperature and the momentum spread
of the condensate in Sec.~\ref{sec_thermal}.

\section{Experimental considerations}

In this section we consider the effects of the physics that our
homogeneous calculations were not able to capture, and estimate
their effect on the damping of the Rabi oscillations.

\label{sec_thermal}

\subsection{Finite temperature}

\begin{figure}[t]
\begin{centering}
\includegraphics[width=7cm]{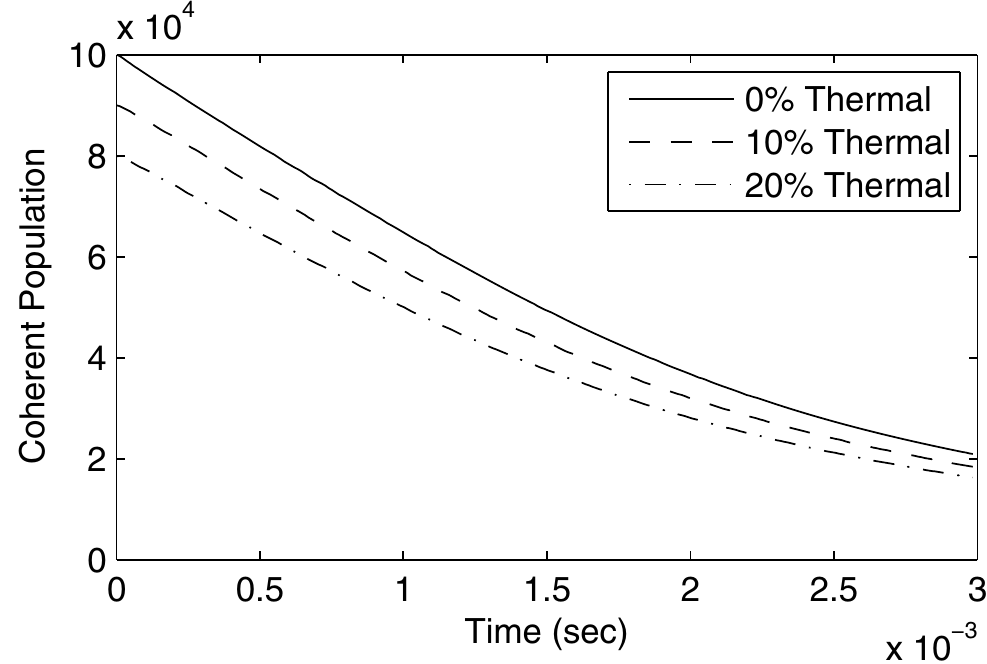}
\caption{The total coherent population of the condensate beginning
at temperatures corresponding to pure condensate, 90\% condensate,
and 80\% condensate. All simulations contain a total population of
$10^5$ atoms.}  \label{thermal3d}
\end{centering}
\end{figure}

The simulations we have presented so far have all begun with initial
states sampled from the zero temperature Wigner distribution for the
condensate. However, if the system is in fact at any non-zero
temperature then any surrounding thermal cloud is able to increase
the rate of depletion by means of Bose-enhancement of the resonant
collisions. We have performed a 3D simulation with the same
parameters as described in Sec.~\ref{sec_dyn_instd}, but with a 10\%
and 20\% thermal fraction surrounding the pure condensate centered
with momentum $+\!\hbar k_L$. Note that the experiments began with
approximately a 20\% thermal fraction.

In Fig.~\ref{thermal3d} we plot condensate fractions as a function
of time. Fermi's golden rule for scattering predicts that the
scattering rate is proportional to the square of the number of
condensate atoms (\textit{cf}. Eq.~(\ref{fermi})). However, such a
calculation does not take into account the population of the modes
to which atoms are scattered into. A closer look at the results in
Fig.~\ref{thermal3d} indicate that the fractional rate of loss of
the coherent component is actually slightly greater in the cases
where a thermal cloud is present, after a time of 200 $\mu$s. The
Bose-enhancement due to the thermal fraction outweighs the reduction
in condensate fraction.

The oscillations are also damped more quickly at higher
temperatures. An oscillating exponential decay yields damping times
of $\tau \approx 2.00$ ms, 1.76 ms, and 1.61 ms for 0\%, 10\% and
20\% thermal fraction respectively.  The presence of a thermal fraction
can partially account for the faster
rates seen in the experiment.

\subsection{Peak-Density Calculation}

Until now our choice of density for the calculations has been the approximate
average density of the condensate in the experiment. Here we repeat the
simulations using instead the peak density of $1.7 \times 10^{20}$ atoms
m$^{-3}$.
We have no \emph{a priori} reason to prefer one over the other.
However, as the scattering will occur quickest in the region of highest
density, one might expect the dynamics in the peak-density region to dominate.

We performed simulations at zero temperature and with a 20\% thermal
fraction, finding damping rates of 1.22 ms and 1.00 ms respectively.
These numbers are closer to the experimental results. In the next
section we combine this with linear dephasing for a full
comparison.

\subsection{Dephasing and Instability}

\begin{table}
\begin{center}
\begin{tabular}{|c|c|c|c|}
  \hline
  \hspace{1mm} \textbf{Description} \hspace{1mm} &
  \hspace{1mm} \textbf{Density} \hspace{1mm} & \hspace{1mm} \textbf{Thermal} \hspace{1mm} & \hspace{1mm} \textbf{Damping} \hspace{1mm}  \\
  & (m$^{-3}$) & \textbf{fraction} & \textbf{time} \\
  \hline
  Linear Theory & - &  - & 954 $\mu$s \\
  \textit{Linear Exp.} & - & - & $400 \pm 200$ $\mu$s \\
  \hline
  Truncated Wigner & $1.0 \times 10^{20}$ & 0\% & 2.00 ms \\
  Truncated Wigner & $1.0 \times 10^{20}$ & 10\% & 1.76 ms \\
  Truncated Wigner & $1.0 \times 10^{20}$ & 20\% & 1.61 ms \\
  Truncated Wigner & $1.7 \times 10^{20}$ & 0\% & 1.22 ms \\
  Truncated Wigner & $1.7 \times 10^{20}$ & 20\% & 1.00 ms \\
  \hline
  GPE in trap & - & - & 1.3 ms \\
  Combined Theory & $1.7 \times 10^{20}$ & 20\% & 560 $\mu$s \\
  \textit{Trapped Exp.} & - & $\sim 20\%$ & $450 \pm 80$ $\mu$s \\
  \hline
\end{tabular}
\end{center}
\caption{The rates of damping in the simulations and experiments.}
\label{tab_rates}
\end{table}

A summary of the results of all our 3D calculations and the
experimental results are shown in Table~\ref{tab_rates}.
We can now combine the results of the linear dephasing and
the simulation of the dynamical instability. In the simplest model for decay
of the Rabi oscillations, we would simply add the two exponential
decay rates by $\tau^{-1} = \tau^{-1}_{\mathrm{dephasing}} +
\tau_{\mathrm{instability}}^{-1}$. We use the dephasing found in the
Gross-Pitaevskii model and instability at the peak density of $1.7
\times 10^{20}$ m$^{-3}$ with a 20\% thermal fraction. This results in a decay
constant of $\tau
\approx 560$~$\mu$s, which is slightly longer than in the experiment.  Given the
uncertainties involved in calculating this result we consider that it is in rather good
agreement.

\section{Conclusions}

\label{sec_conclusion}

In summary, we have described experiments demonstrating the
interaction-induced instability of a BEC at the band-edge of a 1D
optical lattice. We have qualitatively studied the effects of dynamical
instabilities in one-, two- and three-dimensional systems, demonstrating
different regimes of the dynamical instability.  We have attempted to
quantitatively model the dynamics of this system by using the truncated Wigner
method, with good agreement between our experimental and theoretical
results.

Our truncated Wigner simulations produced slightly slower scattering
and damping of Rabi oscillations than observed in the experiment and
showed that the scattering rate is larger for non-zero temperatures.
However, in  our simulations we have neglected the trapping
potential in order find a regime in which the truncated Wigner
technique was accurate. We have investigated some of the
consequences of the trap using simple linear and mean-field calculations, and
found their contribution to damping of Rabi oscillations.

Although the truncated Wigner method has had limitations in the
particular experiments presented here, there are many other experiments to which
it could be applied. In particular, systems of
reduced dimensionality where the condensate is tightly confined in
one or two directions (i.e. waveguides) will be well modeled with
the truncated Wigner method. This will allow quantitative analysis
of non-classical effects in condensates, such as spontaneous
scattering, number squeezing effects and entanglement between atoms
and with light.

\begin{acknowledgments}

AJF and MJD would like to thank Murray Olsen and Ashton Bradley for
useful discussions, and Elena Ostrovskaya for initiating our
interest in thermalization in lattices.  This work was funded by the
Australian Research Council Centre of Excellence for Quantum-Atom Optics
and the Marsden Fund of New Zealand.

\end{acknowledgments}

%\bibliography{matt}

\end{document}